# An End to End Deep Neural Network for Iris Segmentation in Unconstraint Scenarios

Shabab Bazrafkan[1], Shejin Thavalengal[2], and Peter Corcoran[3]

## Abstract

**With the increasing imaging and processing capabilities of today's mobile devices, user authentication using iris biometrics has become feasible. However, as the acquisition conditions become more unconstrained and as image quality is typically lower than dedicated iris acquisition systems, the accurate segmentation of iris regions is crucial for these devices. In this work, an end to end Fully Convolutional Deep Neural Network (FCDNN) design is proposed to perform the iris segmentation task for lower-quality iris images. The network design process is explained in detail, and the resulting network is trained and tuned using several large public iris datasets. A set of methods to generate and augment suitable lower quality iris images from the high-quality public databases are provided. The network is trained on Near InfraRed (NIR) images initially and later tuned on additional datasets derived from visible images. Comprehensive inter-database comparisons are provided together with results from a selection of experiments detailing the effects of different tunings of the network. Finally, the proposed model is compared with SegNet-basic, and a near-optimal tuning of the network is compared to a selection of other state-of-art iris segmentation algorithms. The results show very promising performance from the optimized Deep Neural Networks design when compared with state-of-art techniques applied to the same lower quality datasets.**

## 1 Introduction

Biometric technology has become increasingly integrated into our daily life- from unlocking the smartphone to cash withdrawals from ATMs to shopping in the local supermarket [1]. Various biometric modalities such as face, iris, retina, voice, fingerprints, palm prints, palm geometry are being used in a multitude of applications including law enforcement, border crossing and consumer applications [2], [3]. The iris of the human eye - the annular region between the pupil and sclera - is of particular interest as iris is a biometric modality with high distinctiveness, permanence and performance [4].

The historical evolution of Iris recognition systems can be broadly summarised by a number of key stages, each presenting a new set of unique challenges over earlier implementations of the technology:

(i)   The original proposal for the use of the iris as a biometrics was made by the ophthalmologist Burch in 1936 [5] and the underlying technology to automate iris recognition for practical deployment was proposed and subsequently developed by Daugman [6] during the 1990's. Such

[1] With the Department of Electronic Engineering, College of Engineering, National University of Ireland Galway, University Road, Galway, Ireland. (E-mail: s.bazrafkan1@nuigalway.ie).
[2] With Xperi Galway, Galway, Ireland. (E-mail: shejin.thavalengal@xperi.com).
[3] With the Department of Electronic Engineering, College of Engineering, National University of Ireland Galway, University Road, Galway, Ireland. (E-mail: peter.corcoran@nuigalway.ie).

The research work presented here was funded under the Strategic Partnership Program of Science Foundation Ireland (SFI) and co-funded by SFI and FotoNation Ltd. Project ID: 13/SPP/I2868 on "Next Generation Imaging for Smartphone and Embedded Platforms".

(i) systems acquired iris pattern using a dedicated imaging system that constrains the target eye and employs near-infrared (NIR) imaging.

(ii) Systems supporting the acquisition of iris pattern from mobile persons, in unconstrained acquisition conditions, were developed during the 2000's, with the iris on the move system from Sarnoff being one of the better known of these [7]. This system was designed for deployment in public spaces such as airports, and requires people to walk along a specified path where multiple successive iris images are acquired by a multi-camera systems under controlled lighting conditions.

(iii) Most recently iris recognition has been developed and deployed on handheld devices including smartphones [2]. Such image acquisition is unsupervised and to a large extent unconstrained. This introduces new artifacts that are not found in earlier acquisition environments including unwanted reflections, occlusions, non-frontal iris images, low contrast and partially blurred images. Isolating iris regions accurately in such an acquisition environment has proved to be more challenging and requires improvements to the authentication workflow as will be discussed shortly.

Majority of the existing iris recognition system follows the authentication workflow as (i) image acquisition: an eye image is acquired using a camera, (ii) iris segmentation: eye/iris region is located in this image followed by isolating the region representing iris. (iii) Feature extraction: relevant features which represent the uniqueness of the iris pattern is extracted from the iris region and (iv) similarity of the two iris representation is evaluated by pattern matching techniques.

The work presented in this paper focuses on successful segmentation of non-ideal iris images, an essential element of the authentication workflow if an unacceptably high levels of failed authentications is to be avoided.

## 1.1 Significance of iris segmentation

Iris segmentation involves the detection and isolation the iris region from an eye image. The subsequent feature extraction and pattern matching stages of any authentication workflow rely on the accurate segmentation of the iris and failed segmentations represent the single largest source of error in the iris authentication workflow [8]–[10]. For an accurate segmentation the exact iris boundaries at pupil and sclera have to be obtained, the occluding eyelids have to be detected, and reflections have to be removed, or flagged. Errors at the segmentation stage are propagated to subsequent processing stages [8], [11]. Detailed analysis of the impact of iris segmentation is studied in [8]–[10].

Numerous factors can introduce challenges in accurate iris segmentation [12] even on high-resolution iris systems. Examples include (i) occlusions caused by the anatomical features of the eye; (ii) illumination conditions; (iii) user cooperation; (iv) environmental factors; (v) noise & manufacturing variations in image sensor technology; (vi) nature of the interacting population. These factors apply to all iris acquisition systems.

For mobile devices, in addition to these generic factors, there are additional concerns. Various image quality factors can also affect iris segmentation [13] and these become a limiting factor in consumer devices such as smartphones due to the challenging nature of acquiring suitable high-quality images in a user-friendly smartphone use-case [14]. Hence, an iris segmentation technique which can accurately isolate the iris region in such low-quality consumer images is important for the wider adoption constraint-free consumer iris recognition system.

This work proposes to significantly improve the quality of iris segmentation on lower quality images by introducing an end to end deep neural network model accompanied by an augmentation technique. These improvements should enable improved iris authentication systems for today's mobile devices, encouraging a broader adoption of iris recognition in day-to-day use cases.

## 1.2 Related Literature & Foundation Methods

### 1.2.1 Iris Segmentation

Iris segmentation has widely studied in the literature. A detailed review of iris segmentation literature can be found in [15], [16]. Early work on iris segmentation approximated the pupillary and limbic boundaries as circles [15]. An appropriate circular fitting method is incorporated for modeling these boundaries. Daugman's original work uses an integrodifferential operator for iris segmentation [17]. This integrodifferential operator acts as a circular edge detector which searches over the image domain for the best circle fit. Applying this operator twice, one can obtain the two circular boundaries of iris. After this step, the occluding eyelashes are detected with the help of curvilinear edge detection. There have been several similar techniques for iris segmentation such as the algorithm proposed by Wildes et al.[18], Kong et al. [19], Tisse et al. [20] and Ma et al. [21]. (All of these use circular Hough transform for finding the circles). Another segmentation technique proposed by He et al. [22] uses an Adaboost-cascade iris detector and an elastic model named 'pulling and pushing method'.

Further studies revealed that iris and pupil boundaries are not circular always, and modeling this accurately, improves the iris recognition performance [23]. Daugman's follow up work [23] incorporates active contours or snakes to model the iris accurately. A similar approach proposed by Shah and Ross [24] uses geodesic active contours for accurate iris segmentation. Such techniques were shown to have high segmentation accuracy in good quality images captured using dedicated iris cameras in the NIR region of electromagnetic spectrum. Proenca and Alexandre noted the poor performance of iris segmentation techniques developed for good quality images when applied to non-ideal images [25]. A recent literature survey on non-ideal iris image segmentation can be found in [26]. Among the literature, it is worth to be noted the efforts of mobile iris challenge evaluation (MICHE) to provide a forum for comparative research on the contributions to the mobile iris recognition field [27], [28]. Techniques based on various adaptive filtering and thresholding approaches are shown to be performing well in these non-ideal scenarios [29], [30].

### 1.2.2 Applications of CNNs in Iris Recognition

In the last decade, deep learning techniques became the center of attention and the most successful approach in artificial intelligence and machine vision science. Deep learning based techniques are noted to provide state of the art results in various applications such as object detection [31], face recognition [32], driver monitoring systems [33], etc. In such deep learning based approaches, the input signal (image) is processed by consecutive signal processing units. These units re-orient the input data to the most representative shape considering the target samples. The signal processing units are known as layers which could be convolutional or fully connected. The fully convolutional model, such as the one presented in this work, which uses only convolutional layers. These layers apply filters (known as kernels) to their input while the filter parameters are learned in the training step. In order to get better convergence, several techniques including drop-out [34] and batch normalization [35] are presented in the literature. A detailed introduction to CNNs and its applications can be found in [36].

Recently, deep learning and convolutional neural networks are applied in the domain of iris recognition. [37] proposed the deep features extracted from VGG-Net for iris recognition. Authors in this work skipped iris segmentation step in their framework, and hence it can be considered as a peri-ocular recognition more than iris recognition. [38] proposed a generalizable iris recognition architecture for iris representation. An open source OSIRIS implementation for iris segmentation is used. While authors note high generalisability and cross-sensor performance, the segmentation errors generated by OSIRIS could be affecting the result of this system. Liu et al. proposed DeepIris for heterogeneous iris verification [39]. Also, deep learning based approaches for spoof and contact lens detection can be found in [40][41]

### 1.2.3 CNN for iris segmentation

Li et al. produced two CN based models for iris segmentation [42]- (i) hierarchical convolutional neural network (HCNN) with three blocks of alternative convolutional and pooling layers fed directly in to a fully connected layer; and (ii) multi-scale fully convolutional network (MFCN) which contains six blocks of interconnected alternative Conv and Pool layers fused through a single multiplication layer followed by a Softmax layer. Jalilian and Uhl [43] proposed three types of fully convolutional encoder-decoder networks for iris segmentation. Arslan et al. [44] proposed a two-stage iris segmentation based on CNNs for images captured in visible light. Authors used circular Hough transform to detect rough iris boundary in the first stage. A pre-trained VGG-face model is used in the second stage for the fine adjustment of rough iris boundary obtained in the first stage. In order to overcome the requirement of large labeled data in the approaches mentioned above, Jalilia, Uhl and Kwitt proposed a domain adaption technique for CNN based iris segmentation [45].

### 1.2.4 Foundation Methods

The two primary contribution of this work are (i) a novel iris database augmentation and (ii) a semi parallel deep neural network.

#### *1.2.4.1 Database Augmentation*

Since deep learning approaches need a large number of samples to train a deep network, data augmentation becomes a crucial step in the training process. Database augmentation is the process of adding variation to the samples in order to expand the database and inject uncertainty to the training set which help the network avoid overfitting and also generalizing the results. Also, the augmentation step can introduce more variations into the database and helps the network to generalize its results. The most well-known augmentation techniques include flipping, rotating and adding distortions to the image are widely used in expanding databases. Such techniques are usually used blindly and do not always guarantee any boost in the performance [46].

In [46], authors proposed a smart augmentation method which combines two or more samples of the same class and generates a new sample from that class. This method can give superior results compared to classical augmentation techniques. Unfortunately, this method is only applicable to classification problems, and does not take into account the variations that are needed for a specific task. The other importance of the augmentation is that it gives the ability to manipulate the network results toward a specific condition, for example adding motion blur to the samples can introduce the robustness to the motion blur in the final results.

The data augmentation technique employed in this work is designed for the specific task of iris recognition. A good quality, ISO standard compliant iris image [47] is degraded to make them a good representation of the real-world, low-quality consumer grade iris images. The degradation reduces iris-pupil and iris sclera contrast along with introduction of various noises. Iris map obtained from state of the art iris segmentation techniques are used to aid this process. This particular strategy of augmentation is employed to harness the highly accurate segmentation capabilities of the state of the art iris segmentor. In this way, images which are a reliable representation of the low-quality consumer images can be obtained along with the corresponding iris map for the training of the neural network.

*1.2.4.2   Semi Parallel Deep Neural Network (SPDNN)*

The second contribution of this work is the use of the recently introduced network design method called Semi Parallel Deep Neural Network (SPDNN) [48], [49] for generating iris maps from low quality iris images. In an SPDNN, several deep neural networks are merged into a single model to take advantage of every design. For a specific task, one can design several DNN models each of them having advantages and shortcomings. The SPDNN method gives the possibility of merging these networks in layer level using graph theory calculations. This approach maintains the order of the kernels from the parent networks in the merged network. The convergence and generalization of this method along with various application can be found in [49], [48]. In the present work, the model is trained to generate an iris map from such low-quality images.

## 1.3   Contribution

This work targets the iris segmentation in low-quality consumer images such as the images obtained from a smartphone. An end to end deep neural network model is proposed to isolate iris region from the eye image. The proposed segmentation technique could be used with any existing state of the art feature extraction and matching module without changing the whole authentication workflow. Performance evaluation of the proposed technique shows advantages over recent iris segmentation techniques presented in the literature. There are notably three primary contributions in this work.

1- An improved data augmentation technique optimized to generate diverse low-quality iris images. Such iris images are representative of unconstrained acquisition on a handheld mobile device from multiple established iris research databases.
2- A sophisticated iris segmentation network design derived using Semi Parallel Deep neural Network techniques; Design and optimization methodologies are presented in detail.
3- A detailed evaluation of the presented iris segmentation approach is presented on various publically available databases. The presented method is compared with state of the art techniques in iris segmentation.

In the next section, the database and augmentation technique is explained. The network design and Training is explained in section three followed by results given in section four. The last section explains the numerical results, experiments on tuning, and comparisons to state of the art segmentation methods.

# 2   Databases & Augmentation Methodology

In this work, four datasets are used for training and evaluation. Bath800 [50] and CASIA Thousand [51] have been used in training and testing stages. UBIRIS v2 [52] and MobBio [53] are taking part in tuning and also testing. Bath800 and CAISA thousand has been augmented to represent more Real world consumer grade situations. 70% of the samples have been used for training/tuning, 20% for validation and 10% for testing. The network is trained initially on CASIA Thousand and Bath800 and tested on all databases. For further observations, the original network has been tuned on UBIRIS v2 and MobBio separately and also on a mixed UBIRIS+MobBio database. All the experiments and discussions are given in section 5. Following is introducing databases used in this work, followed by the ground truth generation and augmentation.

**CASIAThousand**

CASIA Thousand is a subset of CASIA Iris v4 database, and it contains 20000 NIR images captured from 1000 individuals. Images are taken using an IKEMB-100 camera which is an IrisKing product. This is a dual-eye camera with a user-friendly interface. The database contains samples with eyeglasses and specular reflections. Images are constrained high quality with high contrast. The resolution is [640×480] for all images. Samples of this database are shown in Figure 1.

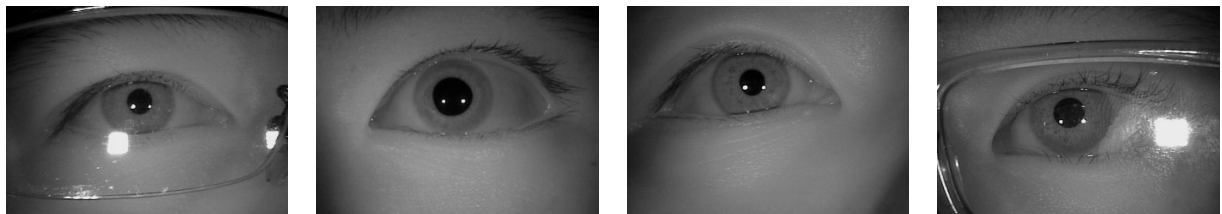
*Figure 1: eye socket samples from CASIA Thousand database*

### Bath800

Bath800 is a high-quality iris database taken under near infrared illumination using a Pentax C-3516 M with 35 mm lens. Images are captured in 20cm distance from the subject. The light source is kept close to the capturing device to reduce the reflections and shadow in the captured images while illuminating the iris texture. Image resolution is [1280×960]. The database is made of 31997 images taken from 800 individuals. The images are high quality and high contrast. Figure 2 shows samples of this database.

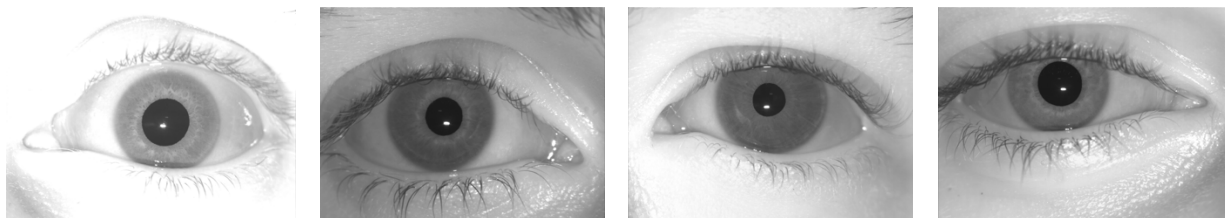
*Figure 2: eye socket samples from Bath800 database*

### UBIRIS v2

UBIRIS v2 database has 11102 images taken in visible wavelength with a Canon EOS 5D which are relatively low-quality images captured in an unconstrained environment. The database is made from 261 participants, 522 Irises. Images include samples in motion, off-axis, occluded, with reflections and glasses, taken in the distance with several realistic lighting conditions. The resolution of the database is 400×300. This database is not used in our training step. The network has been tested on UBIRIS v2 and then tuned on this database for further evaluations described in section 5.Some samples of this database are shown in Figure 3.

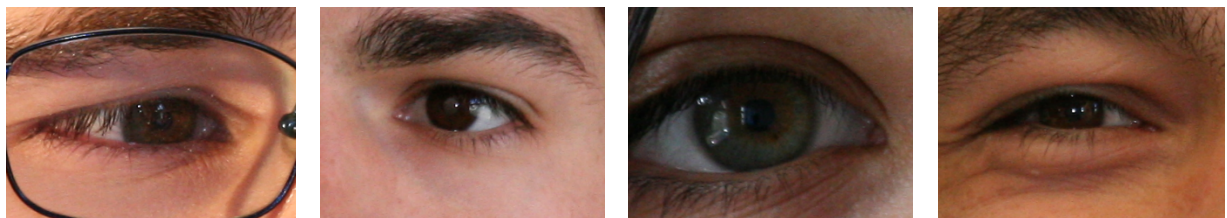
*Figure 3: eye socket samples from UBIRIS database*

### MobBio

MobBio is a multimodal database including face, iris, and voice of 105 volunteers. We used the iris subset of this database in the current work. These images are taken under different lighting conditions including natural and artificial ones. Images are taken in the distance range of 10cm to 50cm of the subject. The device used to capture images was ASUS Transformer Pad TF 300T back camera (TF300T-000128, 8 MP with autofocus). The iris images are taken in several orientations, with different levels of occlusion. 16 images are taken from each individual and cropped and resized to resolution 300×200. This database is highly unconstrained and one of the most challenging sets in iris segmentation/recognition. Some samples of this database are shown in Figure 4.

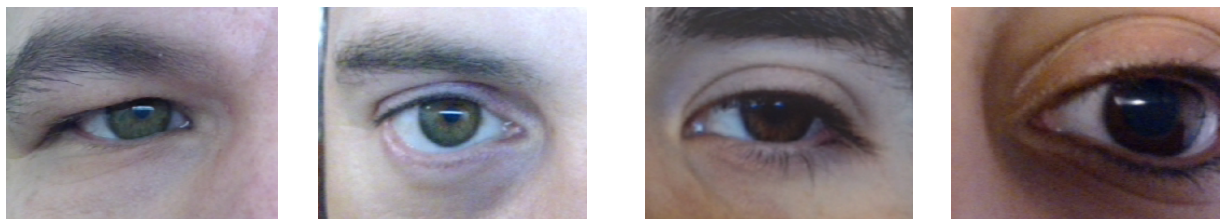
Figure 4: eye socket samples from MobBio database

## 2.1 Ground Truth generation
### 2.1.1 Bath800 and CASIA Thousand
Neither of Bath800 and CASIA Thousand databases are provided with ground truth segmentation. As mentioned before these databases contain very high-quality images taken in highly constrained conditions, having high resolution, high contrast, limited shadow, and low noise level. This gives us the opportunity to be able to apply industry standard segmentation algorithms and get very high accuracy segmentation results. In this work, the segmentation from high-quality images obtained using a commercial iris segmentation solution (MIRLIN [54]) is considered as the ground truth for training stage.

It can be noted that, any commercial, high performing segmentation technique could be used here as these high quality images could be segmented accurately using such commercial systems. This specific choice of segmentor used in this work is based on its availability and its performance on large scale iris evaluations [55] Some segmentation examples are given in Figure 5 and Figure 6. The low resolution segmentation for Bath800 and CASIA Thousand is publicly available[4].

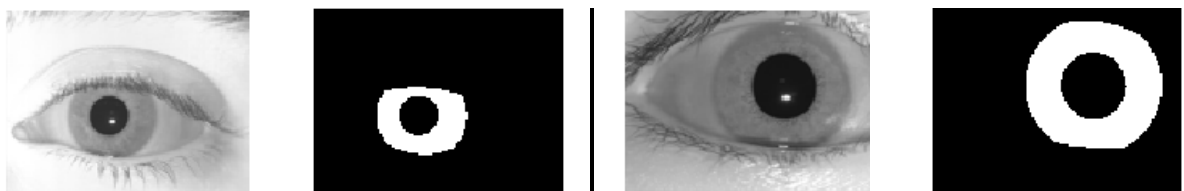
Figure 5: Bath800 automatic segmentation results

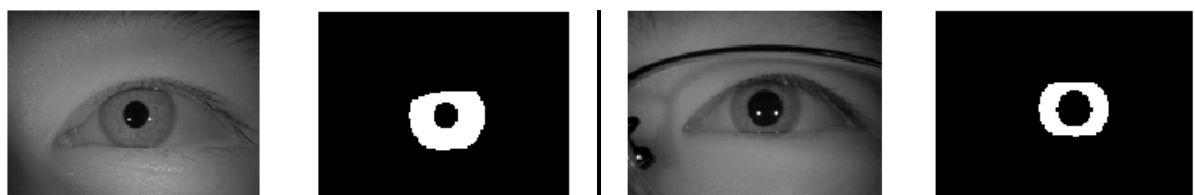
Figure 6: CASIA Thousand automatic segmentation results

### 2.1.2 UBIRIS and MobBio
The manual segmentation of UBIRIS is available in IRISSEG-EP database [11] generated by WaveLab [56]. In this database, the iris map is determined by eclipses for the iris boundary and polynomials for eyelids. The inner and outer circles of the iris have been identified by at least 5 points, and ellipses have been fit into the points using the least square method. The upper and lower eyelids have been identified with at least 3 points and second order polynomials have been fit by the mean square method. The ground truth generation for UBIRIS is not completed. Only segmentations for 2250 images from 50 individuals are given in IRISSEG-EP. Some samples for UBIRIS segmentation is shown in Figure 7.

---
[4] https://Goo.gl/JVkSyG

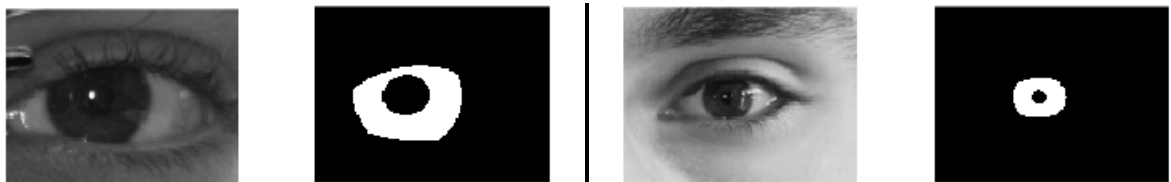

*Figure 7: UBIRIS manual segmentation samples in IRISSEG-EP*

The manual segmentation for the MobBio database is included in IRISSEG-CC database [11] generated by ISLAB [57]. The whole MobBio dataset has been segmented in IRISSEG-CC. In this database, the inner and out iris circumferences and also the upper and lower eyelids are identified by circles given the radius and center. Some samples for MobBio segmentation are given in Figure 8.

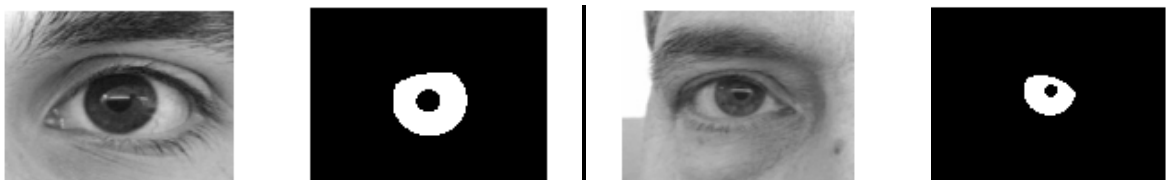

*Figure 8: MobBio manual segmentation samples in IRISSEG-CC*

## 2.2 Data Augmentation

In this work two high-quality databases, Bath800 and CASIA Thousand have been used in training stage. In this section, we are going to describe the augmentations which have been applied to the high-quality iris images, in order to simulate the real-life environments in iris recognition task.

In order to find the best augmentations for the iris images, precise observations have been done on low-quality wild iris images. The difference between a high quality constrained iris images and consumer grade images depend on five different independent factors: 1- eye socket resolution, 2- image contrast, 3- shadows in the image, 4- image blurring, 5- noise level [14], [58]. In our observations, the noise level was low in wild images. Noise is a well-studied phenomenon and image de-noising can be done outside the network and also note that introducing high frequency noise into the dataset trains a low-pass filter inside the network; apply de-noising outside the network gives a higher chance to use the whole network potential to perform the segmentation task. In addition, introducing Gaussian noise into the dataset will cause underfitting as explained in [59]. Therefore in this work we mostly focused on the first four factors.

These are next discussed in turn with details of the augmentation approaches taken for each independent factor. The code for augmenting the database is also available[5].

## 2.2.1 Eye socket resolution

The resolution of the eye socket plays an essential role in how much information one can extract from the image. In fact, while dealing with the front camera in a mobile phone, the resolution of the camera is lower than the rear cameras. For example, Bazrafkan, Kar and Corcoran [60] observed that, the number of the pixels in the iris region for an image taken by a 5MP front camera of a typical cell phone from the 45cm distance was just 30 pixels. This low resolution will make iris segmentation/recognition task more vulnerable to other effects like hand jitter and motion blurring.

In order to simulate the low resolution scenario, the high-quality eye socket images and their corresponding ground truth have been resized using bilinear interpolation into smaller images [128×96]. This can help the deep network to train faster as well.

---
[5] https://github.com/C3Imaging/Deep-Learning-Techniques/blob/Iris_SegNet/DBaugmentation/DBaug.m

### 2.2.2 Image Contrast

The iris images acquired by handheld devices with poor optical properties are significantly different from the high-quality, high-resolution NIR images obtained from a constrained, acquisition scenario. The contrast inside and outside the iris region is different for these two type of images. In fact in the low-quality image set, the region inside the iris was darker than the same region for the high-quality images, and the contrast was lower as well. There was no specific brightness quality for the regions outside the iris in low-quality images. They could be customarily exposed or strongly bright or very dark, because of the unconstraint environments. Again in the low-quality images, the regions outside the iris were suffering from the low amount of contrast.

In order to apply this transformation to the high-quality images, we defined two steps; one targeting outside of the iris, other for pixels inside the iris region.

1- Outside iris region: As explained before the pixels outside the iris region are suffering from low contrast intensities. It could be bright, normal exposure or dark. Histogram mapping technique is used to reduce the contrast. The histogram transformation is given by

$$y = norm\big(\tanh(3\times(x/255 - 0.5) + \mathcal{U}(-0.3, 0.3))\big) \times 255 \quad (1)$$

where $x$ is the input intensity in the range [0,255], $y$ is the output intensity in the same range, $\mathcal{U}(a, b)$ is the Uniform distribution between $a$ and $b$, and the $norm$ function normalize the output between 0 and 1. The Uniform distribution injects an amount of uncertainty into the mapping which helps the generalization of the model.

The mean and standard deviation of the histogram mapping curve is shown in Figure 9. As one can observe, a symmetrical mapping is selected for the outside region in order to reduce the contrast without changing the brightness significantly.

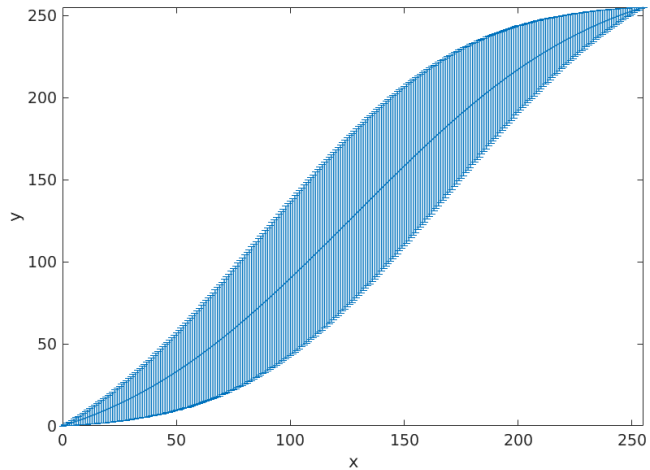

*Figure 9: The histogram mapping for the outside region of the iris.*

2- Inside iris: As noted before, it is observed that the inside iris region possesses low contrast and low intensity. The histogram mapping technique is used to shift the pixel intensities into darker regions and reduce the contrast at the same time. The following histogram mapping is used for inside iris region.

$$y = norm\big(\tanh(3\times(x/255 - 0.5) - \mathcal{U}(0, 0.8))\big) \times 255 \quad (2)$$

where $x$ is the input intensity in the range [0,255], $y$ is the output intensity in the same range, $\mathcal{U}(a,b)$ is the Uniform distribution between $a$ and $b$, and the $norm$ function normalize the output between 0 and 1. The Uniform distribution introduces uncertainty to the mapping which helps the network to generalize the solution. The mean and standard deviation of the histogram curve is shown in Figure 10. You can see that the transformation is darkening the area while decreasing the contrast at the same time.

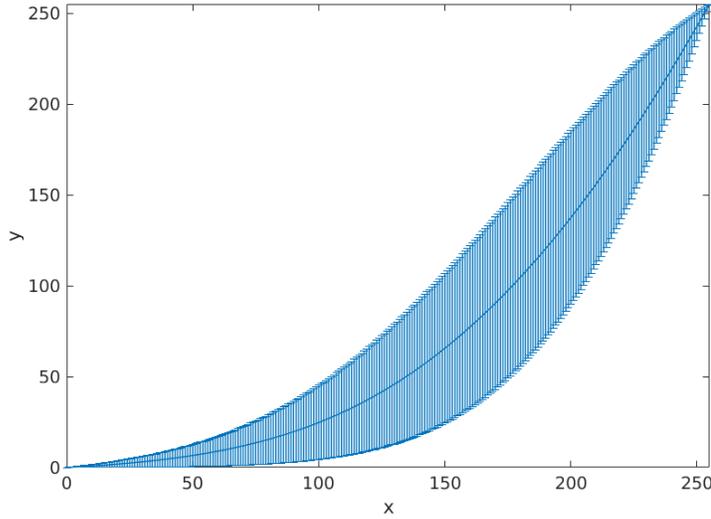

Figure 10: The histogram mapping for the region inside the iris.

These histogram mappings are applied to the iris image. It reduces the contrast for the regions outside iris and darkens the iris region. Figure 11 shows an example of this step.

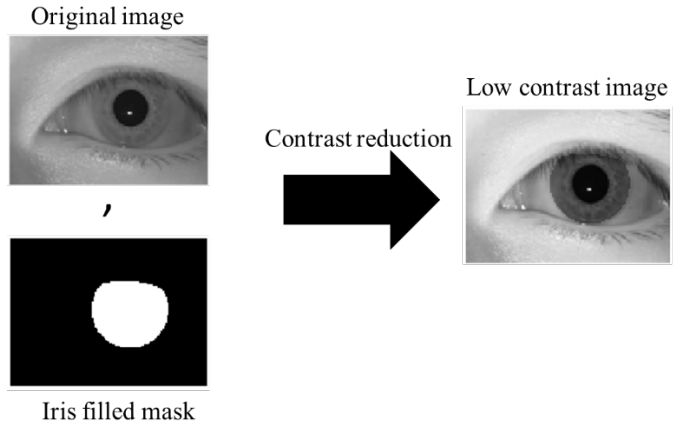

Figure 11: For inside the iris region, the contrast is reduced, and the region is getting darker. The outside of iris is just altered by decreasing the contrast.

### 2.2.3 Shadows in the image

Low quality unconstrained iris images are profoundly altered by the direction of the illumination, in order to be able to train a model which is robust to this effect one needs to enter shadowing into the augmentation process. In this work, shadowing is carried out by multiplying the image columns by the following function.

$$y = norm\left(\tanh\left(2 \times randSign \times (x - 0.5 + \mathcal{U}(-0.3, 0.3))\right)\right) + \mathcal{U}(0, 0.1) \qquad (3)$$

Where $x$ is the dummy variable for image column number and $y$ is the coefficient for intensity, $\mathcal{U}(a,b)$ is the Uniform distribution between $a$ and $b$, and the $norm$ function normalize the output between 0 and 1 and $randSign$ generates a random coefficient in the set $\{-1,1\}$. The Uniform distribution adds an amount of uncertainty to the augmentation process which increases the generalization of the solution. The mean and standard deviation of this function is shown in Figure 12 you can see that based on the value of the $randSign$ function, the shadowing direction is changed.

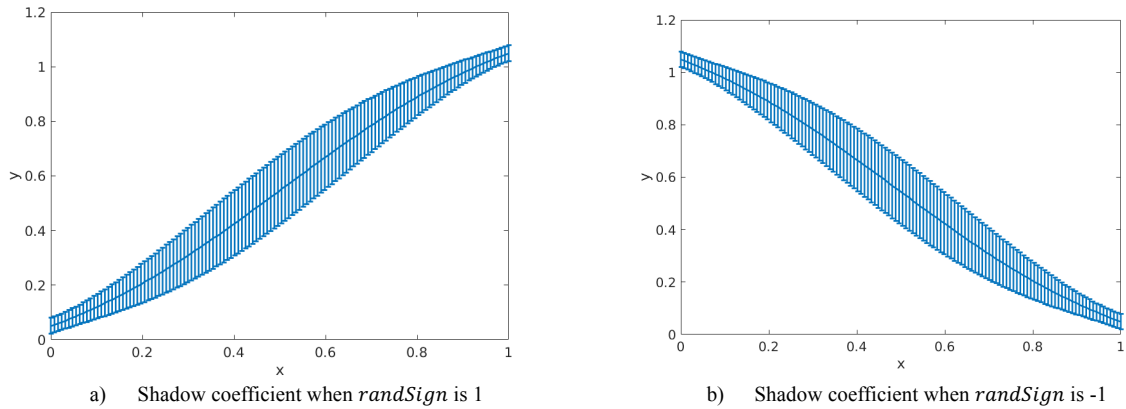

a) Shadow coefficient when $randSign$ is 1  b) Shadow coefficient when $randSign$ is -1

Figure 12: Mean and standard deviation for shadow coefficients

An example of shadowing is given in Figure 13.

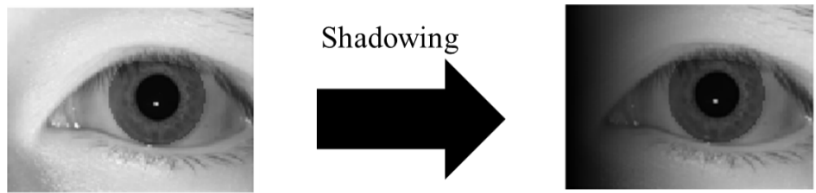

Figure 13: Shadowing applied to low contrast image

### 2.2.4 Image blurring:

Bokeh effect caused by camera focus [61], the hand jitter and unwanted head and hand movements are highly degrading the iris image quality in handheld devices. It became significant when this effect is accompanied with low image resolution and poor optical properties of the device. Such a scenario will make the iris segmentation task challenging. One of the main challenge in iris segmentation in this scenario is the blurring of the iris edges which affects the edge detection task. In majority of the iris segmentation methods, the gradient of the pixel intensity is used to find the iris region and the blurring is highly altering the gradient quality. The deep neural network is able to solve this problem if enough variation of motion blurring is provided in the dataset. In order to include this effect in the training set, shadowed image is passed through a motion filter. A motion filter is a pre-defined filter mimicking the camera motion which accepts the sharp image, the number of pixels (indicating the power of the motion or speed of the camera) and a direction (the direction of the camera motion) and gives back the blurred image. In this work the shadowed image is passed through is motion blur filter applying the linear camera motion by $\mathcal{U}(5,10)$ pixels in the direction $\mathcal{U}(-\pi, \pi)$, where $\mathcal{U}(a,b)$ is the Uniform distribution between $a$ and $b$. The uniform distribution for number of pixels and motion direction is introducing an amount of uncertainty into the database which lead to a more generalized solution. The final image after applying motion blur is shown in Figure 14.

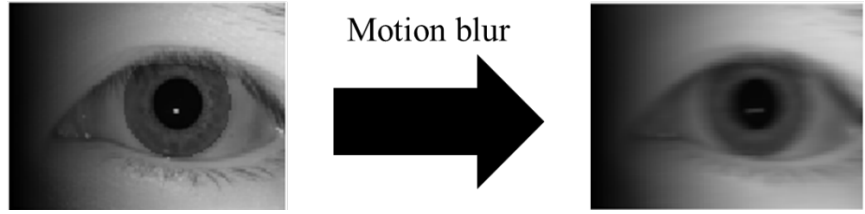
*Figure 14: applying motion blur in a random direction to the low contrast shadowed image.*

All the samples in Bath800 and CASIA Thousand databases are degraded using this augmentation step. Some examples of the low-quality samples and their corresponding ground truth is given in Figure 15.

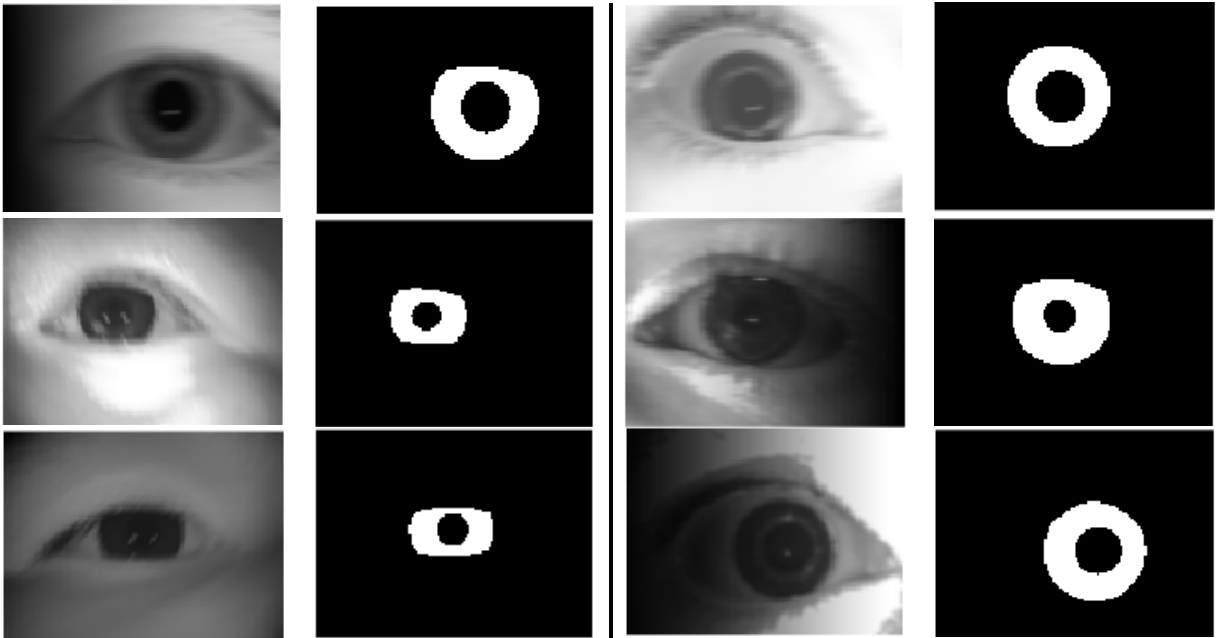
*Figure 15: augmented samples and their corresponding segmentation map.*

# 3 Network Design and Training

## 3.1 Network Design

Deep neural networks are capable of solving highly nonlinear and challenging problems. In this work, four different end to end fully convolutional deep neural networks have been proposed to perform the iris segmentation on low quality images. These networks are merged using SPDNN method and the number of the channels in each layer is selected in a way that the number of the parameters in the proposed network is similar to the SegNet-basic. The network design and calculating the number of channels in each layer are explained in detail in Appendices A and B respectively. The SPDNN method is merging several deep neural networks in the layer level using graph theory calculation and graph contraction. This approach is preserving the order of the layers from the parent networks. The convergence and generalization of SPDNN is discussed in [49] and other applications of this method is given in [48].

The parent networks used in this work are fully convolutional networks with different depths and kernel size each designed to extract different levels of details. The network after merging these networks is shown in Figure 41. The model looks like a U-net [62] with the difference that there is no pooling applied in our proposed network. The number of the channels in each layer is determined in Appendix B. The calculations guarantee that the number of the parameters in the presented method is similar to SegNet-Basic proposed in [63]. Having the same number of parameters helps to obtain a fair comparisons between SegNet-Basic and proposed model.

## 3.2 Training

The proposed network is an end to end design which means that it accepts the eye socket image and gives the iris map. The network has been trained using lasagna library [64] on the top of the theano library [65] in python. The loss function used in our work is the mean binary cross-entropy between the output and target given by

$$L = \frac{1}{M \times N \times B}\sum_{k=1}^{B}\sum_{j=1}^{N}\sum_{i=1}^{M} t_{ij}\log(p_{ij}) - (1 - t_{ij})log(1 - p_{ij}) \qquad (4)$$

Wherein $t_{ij}$ is the value of pixel $(i,j)$ in the target image, $p_{ij}$ is the value of pixel $(i,j)$ in the output image for the image of the size $M \times N$ and $B$ is the batch size. The stochastic gradient descent with momentum has been used to update the network parameters. The momentum term prevents the gradient descent to stick in the local minimums, and also speeds up the convergence. In this approach, the gradient decent uses the update value of the previous iteration as the momentum in the current iteration. Suppose the loss function is $L(w)$ where $w$, is the set of network parameters. The stochastic gradient method with momentum is given by

$$w := w - \eta \nabla L(w) + \alpha \Delta w \qquad (5)$$

wherein $\Delta w$, is the update in the previous iteration, $\nabla L(w)$ is the gradient value in the current iteration, $\eta$ is the learning rate and $\alpha$ is the momentum. In our training experiments, the learning rate and momentum are set to 0.001 and 0.9 respectively. The training method and learning parameters in training the proposed network and SegNet-basic are same.

The network is trained on an augmented version of Bath800 and CASIA1000 originally. Some experiments have been conducted on this original network given in sections 5.2 and 5.3. These databases are NIR databases. In order to provide a network, segmenting visible images, the original network has been tuned on UBIRIS and MobBio databases. The same training method has been used in the tuning stage while learning rate and momentum are set to 0.001 and 0.9 respectively. The train/tune has been done for 1000 epochs. More details on tuning results is given in appendices C and D.

# 4 Results

In the test step, each eye socket image is given to the trained/tuned network, and the forward propagation is performed for this input. In the training stage, the output of the network is forced to converge to the iris segmentation map which is a binary image. The output of the network is a grayscale segmentation map, and the binary map is produced by thresholding technique, i.e., the values bigger than a threshold are shifted to 1 and the others to 0. The threshold value 0.45 has been used in our experiments. The output of the proposed model for different databases are shown in Figure 16 to Figure 19.

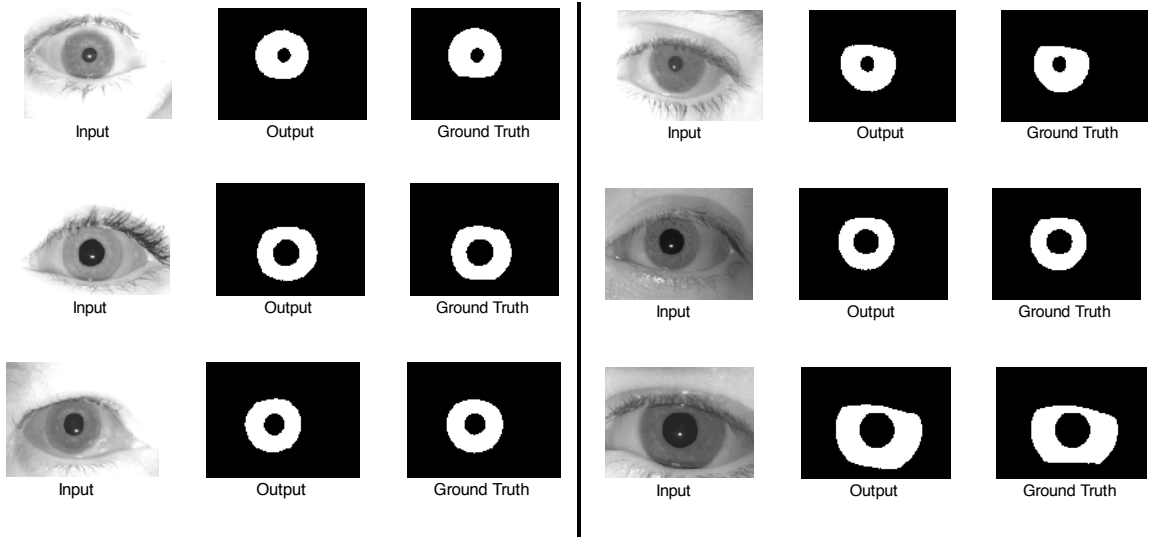

*Figure 16: output of the network for Bath800 test set. The results show high-quality output in this database.*

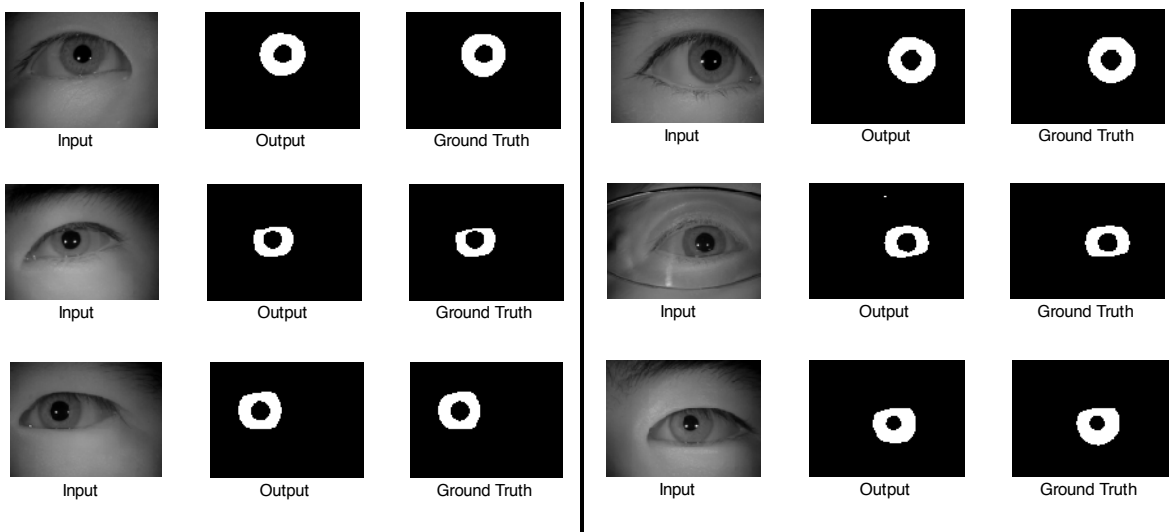

*Figure 17: output of the network for CASIA1000 test set. The results show high-quality output in this database.*

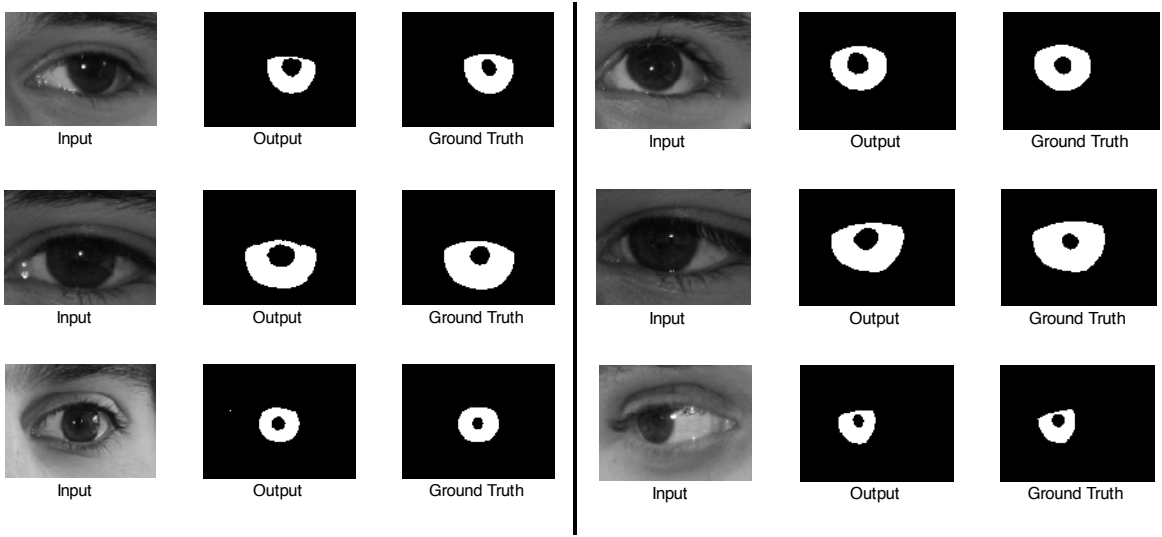

*Figure 18: output of the network for UBIRIS test set.*

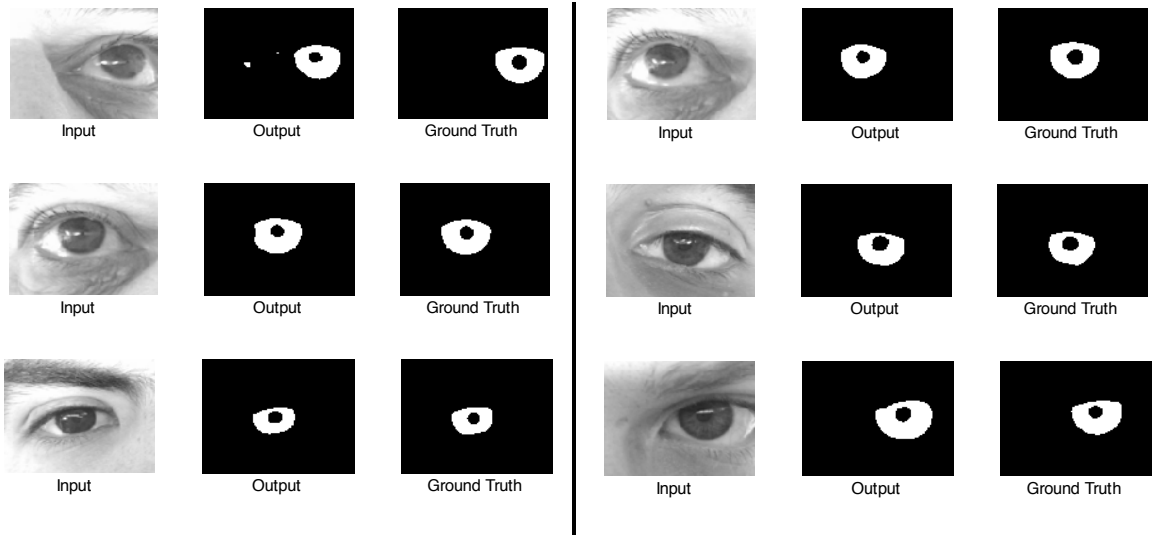

*Figure 19: output of the network for MobBio test set.*

Figure 16 and Figure 17, show the high-quality output for Bath800 and CASIA1000 databases. These datasets are high quality constrained NIR sets, and their images follow a specific distribution which makes it easier for the DNN to perform the segmentation task. Figure 18 and Figure 19, show the output of the proposed network for more difficult unconstrained UBIRIS and MobBio databases. These two figures show the results of the network tuned on these databases. The results are not as good as Bath800 and CASIA1000, but one should note that these datasets are quite challenging and difficult to segment. The numerical results are given in the following section.

## 5  Evaluations

Several metrics have been used to evaluate the network and investigate the tuning effect on the segmentation results. These metrics are presented in Table 1. In all equations True Positive is abbreviated as TP, True Negative as TN, False Positive as FP and False Negative as FN. Letter P stands for the number of all Positive cases which is equal to TP+FN and N is the total number of negative cases equals to FP+TN.

*Table 1: Metrics used in the evaluation section*

| Measure | Description |
|---|---|
| Accuracy | Accuracy is one of the most representative metrics in evaluating binary classifiers. Accuracy represents the ratio of all true results divided by the number of all samples given by $$Accuracy = \frac{TP + TN}{P + N}$$ |
| Sensitivity or True Positive Rate (TPR) | This measure indicates the ability of the model to recall true positive over all positive samples. i.e., a model with high sensitivity can rule out negative samples more efficiently. Sensitivity is given by $$Sensitivity = \frac{TP}{TP + FN} = \frac{TP}{P}$$ |
| Specificity or True Negative Rate | This measure indicates the ability of the model to recall true negative over all negative samples. i.e., a model with high specificity can find positive samples more efficiently. Specificity is given by $$Specificity = \frac{TN}{FP + TN} = \frac{TN}{N}$$ |
| Precision or Positive Predictive Value (PPV): | Precision is the number of the true positive divided by all the positive cases the model gives. i.e., it is the probability that the positive output is true positive in the space of all positive outcomes, which is given by $$Precision = \frac{TP}{TP + FP}$$ |
| False Discovery Rate (FDR) | Is the complement of precision, which is the probability that the positive output is a false positive. FDR is given by $$FDR = 1 - precision = \frac{FP}{TP + FP}$$ |
| Negative Prediction Value (NPV) | NPV is the number of the true negative divided by all the negative cases the model gives. i.e., it is the probability that the negative output is true negative in the space of all negative outcomes, which is given by $$NPV = \frac{TN}{TN + FN}$$ |
| F1 score | Is the harmonic average of precision and sensitivity. This measures the ability of the model to recall true positive cases and at the same time not missing positive cases. F1 score is given by $$F1\ score = \frac{2TP}{2TP + FP + FN}$$ |
| Matthew Correlation Coefficient (MCC) | Is a metric measuring the quality of binary classifiers. The critical property of MCC is its independence of the size of each class. MCC is using false and true positive and negative samples to compute a metric representing the classifier's quality. It varies in the range (-1,1) wherein 1 indicates the perfect model, 0 shows the output is random compared to the target and -1 declared that all output values are the inverse of the target value. MCC is given by $$MCC = \frac{TP \times TN - FP \times FN}{\sqrt{(TP + FP)(TP + FN)(TN + FP)(TN + FN)}}$$ |
| Informedness | Is a metric showing the probability of informed decision given by the model. It ranges from -1 to 1; where 0 shows a random decision and 1 indicates no false outputs. -One shows that all outcomes are false. Informedness is given by $$Informedness = \frac{TP}{TP + FN} + \frac{TN}{TN + FP} - 1$$ |
| Markedness | Is a measure of the information content and information value of the model's output [66]. Markedness is given by $$Markedness = \frac{TP}{TP + FP} + \frac{TN}{TN + FN} - 1$$ |
| False Positive Rate (FPR) | Is a metric which calculates the probability that the model will make a mistake in returning a negative decision in the space of all negative samples. FPR is given by $$FPR = \frac{FP}{FP + TN}$$ |
| False Negative Rate (FNR) | Is a metric which calculates the probability that the model will make a mistake in returning a positive decision in the space of all positive samples. FNR is given by $$FNR = \frac{FN}{FN + TP}$$ |

## 5.1 Experimental results

Five experiments have been conducted to investigate the performance of the proposed network and the effect of the tuning on the results. These experiments are as follows:

1- Test on the original network: The proposed network is initially trained on the augmented version of the Bath800 and CASIA1000 databases. The first experiment compares the output of this network for different databases. The test set of Bath800 and CASIA1000 and all the samples of UBIRIS and MobBio are used in the test stage. Section 5.2 discusses this experiment in detail.
2- Comparison with SegNet-Basic: This experiment discusses the results of presented network compared the SegNet basic. Training and testing for SegNet-Basic is done similar to the proposed network. This experiment is presented in section 5.3.
3- Tuning; Network experiment: In this experiment the original network trained on the augmented version of Bath800 and CASIA1000 is tuned on UBIRIS and MobBio individually and also on a mixture of these two databases. In this way, the effectiveness of each database in boosting the performance is investigated. The network tuned on each database is tested on all databases. The results and discussions of this experiment is presented in appendix C.
4- Tuning; Database experiment: This experiment is looking at the results of previous experiment based on each database. There are four networks trained and tuned which are presented in experiment 1 and 3 as follows: i) Initially trained on Bath800 and CASIA1000. ii) Tuned on UBIRIS. iii) Tuned on MobBio. iv) Tuned on UBIRIS+MobBio. The output of each of these networks for each database and also the average performance is investigated in this experiment. Appendix D is presenting this experiment in more detail.
5- Comparison to state of the art: In this experiment, the best results of the proposed method is compared with other methods in the literature. The numerical results are presented in section 5.4.

In all experiments, $\mu$ stands for the average value for the given measure and $\sigma$ is its standard deviation over all outputs.

## 5.2 Test on the original network

In this experiment, the proposed trained network is tested over four databases (Bath800, CASIA1000, UBIRIS, and MobBio). The reason for adding two more databases in the testing procedure is to observe the ability of the network in generalizing over other databases. One of the main concerns in DNN community is to be able to generalize the trained network to wild environments. This is happening since the majority of Machine learning schemes are sharing the same database in train and test stage. The neural networks learn the distribution of the data for the given database, and since the test set follows the same distribution, it gives promising results. [67] discusses this problem in more detail. Since the network is trained on a merged version of the Bath800 and CASIA1000; only the test sub-set of these databases has been used in testing stage. However, the network never saw the UBIRIS and MobBio set before. Therefore all samples of these databases have been used for testing. The numerical results are shown in Table 2 and Table 3. From this experiment, we can see that the network gives better results for bath800 and CASIA1000 which is expected.

*Table 2: Testing on the original network. Metrics measured for different databases. Green means higher performance and red declares lower quality results. A Higher value of $\mu$ and lower value for $\sigma$ is desirable.*

|  |  | Bath800 | CASIA1000 | UBIRIS | MobBio |
|---|---|---|---|---|---|
| Accuracy | $\mu$ | 98.55% | 99.71% | 97.82% | 96.12% |
|  | $\sigma$ | 1.43% | 0.33% | 1.49% | 3.16% |
| Sensitivity | $\mu$ | 96.03% | 97.96% | 74.29% | 65.79% |
|  | $\sigma$ | 4.76% | 2.95% | 20.12% | 23.94% |
| Specificity | $\mu$ | 99.10% | 99.82% | 99.25% | 97.96% |
|  | $\sigma$ | 1.07% | 0.20% | 1.08% | 2.19% |
| Precision | $\mu$ | 96.05% | 97.13% | 85.65% | 68.71% |
|  | $\sigma$ | 4.46% | 3.10% | 19.94% | 27.88% |
| NPV | $\mu$ | 99.05% | 99.87% | 98.40% | 97.91% |
|  | $\sigma$ | 1.49% | 0.28% | 1.20% | 1.73% |
| F1-Score | $\mu$ | 95.93% | 97.50% | 78.08% | 65.96% |
|  | $\sigma$ | 3.88% | 2.51% | 19.79% | 25.06% |
| MCC | $\mu$ | 0.951 | 0.9737 | 0.7792 | 0.647 |
|  | $\sigma$ | 0.0421 | 0.025 | 0.1909 | 0.2611 |
| informedness | $\mu$ | 0.9514 | 0.9779 | 0.7354 | 0.6375 |
|  | $\sigma$ | 0.0486 | 0.0297 | 0.2028 | 0.2519 |
| markedness | $\mu$ | 0.951 | 0.97 | 0.8406 | 0.6663 |
|  | $\sigma$ | 0.0468 | 0.0312 | 0.1996 | 0.2891 |

Table 3: Testing on the original network. Metrics measured for different databases. Green means higher performance and red declares lower quality results. A lower value of μ and σ is desirable.

|     |   | Bath800 | CASIA1000 | UBIRIS | MobBio |
|-----|---|---------|-----------|--------|--------|
| FPR | μ | 0.89%   | 0.17%     | 1.85%  | 2.03%  |
|     | σ | 1.07%   | 0.20%     | 2.03%  | 2.19%  |
| FNR | μ | 3.96%   | 2.03%     | 16.40% | 34.20% |
|     | σ | 4.76%   | 2.95%     | 16.22% | 23.94% |
| FDR | μ | 3.94%   | 2.86%     | 24.25% | 31.28% |
|     | σ | 4.46%   | 3.10%     | 22.01% | 27.88% |

The network has already observed these two databases in training stage and learned their distribution which justifies the higher performance on these databases. Having a lower value of sensitivity and precision in UBIRIS and MobBio databases declares the amount of uncertainty of the model in giving back the positive cases. Moreover, the high value of specificity and NPV shows that the trained model was able to rule out non-iris pixels in all databases. The low value of FPR also shows the power of the network in detecting non-iris pixels and the high value of FNR declares that the network is weaker in returning iris-pixels in UBIRIS and MobBio than Bath800 and CASIA1000. The value of F1-score, MCC, Informedness, and Markedness is high for Bath800 and CASIA1000 which indicate the ability of the network to produce consistent segmentations both in finding iris and non-iris pixels for these databases. The same measures return average values for UBIRIS database. This means that the network is generalized for semi-wild environments. Moreover, the low value for MobBio indicates that the network is not much reliable to work in wild environments. Moreover, also the higher amount of FDR shows the high probability of the network in returning false positive for MobBio. In general, we can say that the presented network is reliable in returning non-iris pixels in challenging wild scenarios. Note that both UBIRIS and MobBio are visible databases and that MobBio is a very challenging database. In section 5.4.2 the presented network is compared to other methods on the MobBio database.

### 5.3  Comparison with SegNet-Basic

SegNet [63] is one of the most successful DNN approaches in semantic segmentation. SegNet-basic is the small counterpart of the original SegNet. As explained in Appendix B, our proposed architecture contains almost the same number of parameters as SegNet-basic. This gives us the opportunity to conduct fair comparisons between two models. We trained SegNet-basic on the same data with same hyper-parameters as our proposed model. Table 4 and Table 5 show the results for SegNet-basic tested on four databases Bath800, CASIA1000, UBIRIS and MobBio. Note that the network is tested on the test set of Bath800 and CASIA1000 and all samples of UBIRIS and MobBio.

Table 4: SegNet-basic. Metrics measured for different databases. Green means better quality and red declares lower quality results. A Higher value of μ and lower value for σ is desirable.

|  |  | Bath800 | CASIA1000 | UBIRIS | MobBio |
|---|---|---|---|---|---|
| Accuracy | μ | 97.84% | 99.36% | 94.99% | 94.02% |
|  | σ | 1.68% | 0.46% | 2.95% | 3.73% |
| Sensitivity | μ | 94.20% | 96.22% | 85.41% | 80.67% |
|  | σ | 6.01% | 4.98% | 15.70% | 21.58% |
| Specificity | μ | 98.60% | 99.55% | 95.71% | 94.95% |
|  | σ | 1.41% | 0.30% | 3.01% | 3.34% |
| Precision | μ | 94.04% | 93.01% | 58.60% | 51.93% |
|  | σ | 5.58% | 4.92% | 23.84% | 21.01% |
| NPV | μ | 98.66% | 99.76% | 98.90% | 98.67% |
|  | σ | 1.84% | 0.38% | 1.27% | 1.71% |
| F1-Score | μ | 93.91% | 94.45% | 66.59% | 61.10% |
|  | σ | 4.42% | 3.99% | 20.25% | 20.10% |
| MCC | μ | 0.927 | 0.942 | 0.6694 | 0.6092 |
|  | σ | 0.048 | 0.0395 | 0.1873 | 0.2052 |
| informedness | μ | 0.928 | 0.9578 | 0.8113 | 0.7562 |
|  | σ | 0.0587 | 0.0497 | 0.1622 | 0.2264 |
| markedness | μ | 0.927 | 0.9277 | 0.575 | 0.506 |
|  | σ | 0.0561 | 0.0494 | 0.2355 | 0.216 |

Table 5: SegNet-basic. Metrics measured for different databases. Green means better quality and red declares lower quality results. Lower value of μ and σ is desirable.

|  |  | Bath800 | CASIA1000 | UBIRIS | MobBio |
|---|---|---|---|---|---|
| FPR | μ | 1.39% | 0.44% | 4.28% | 5.04% |
|  | σ | 1.41% | 0.30% | 3.01% | 3.34% |
| FNR | μ | 5.79% | 3.77% | 14.58% | 19.32% |
|  | σ | 6.01% | 4.98% | 15.70% | 21.58% |
| FDR | μ | 5.95% | 6.98% | 41.40% | 48.06% |
|  | σ | 5.58% | 4.92% | 23.84% | 21.01% |

Results show SegNet-basic has considerably better performance on Bath800 and CASIA1000 than other two databases; which is expectable since the network is trained on former databases. In the following subsections, these results are compared to presented network in order to find the advantages and shortcomings of each design.

### 5.3.1 Comparing results on Bath800 and CASIA1000

Since both networks are trained on Bath800 and CASIA1000 databases, the numerical test results show the capability of each design in capturing the probability distribution of the training set. Figure 20 to Figure 23 illustrate the comparisons between the proposed method and SegNet-basic over Bath800 and CASIA100. These figures show the mean value for each metric.

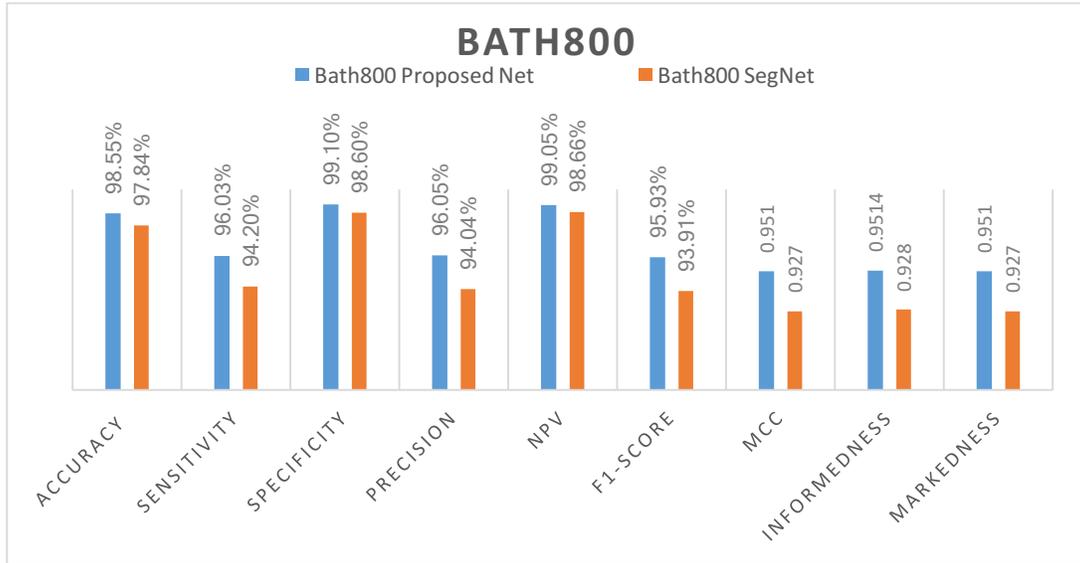

*Figure 20: Comparisons between the presented network and SegNet-basic on Bath800. A Higher value indicates better performance.*

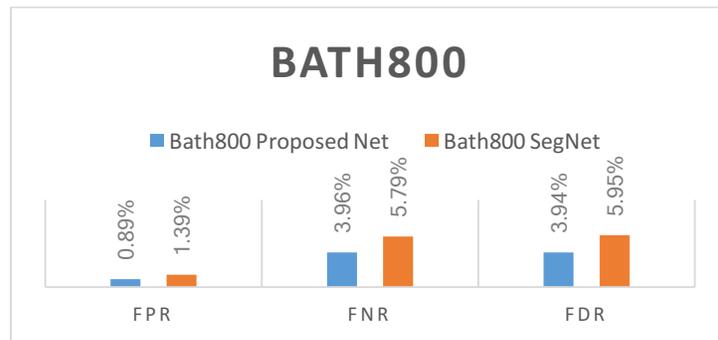

*Figure 21: Comparisons between the presented network and SegNet-basic on Bath800. Lower value indicates better performance.*

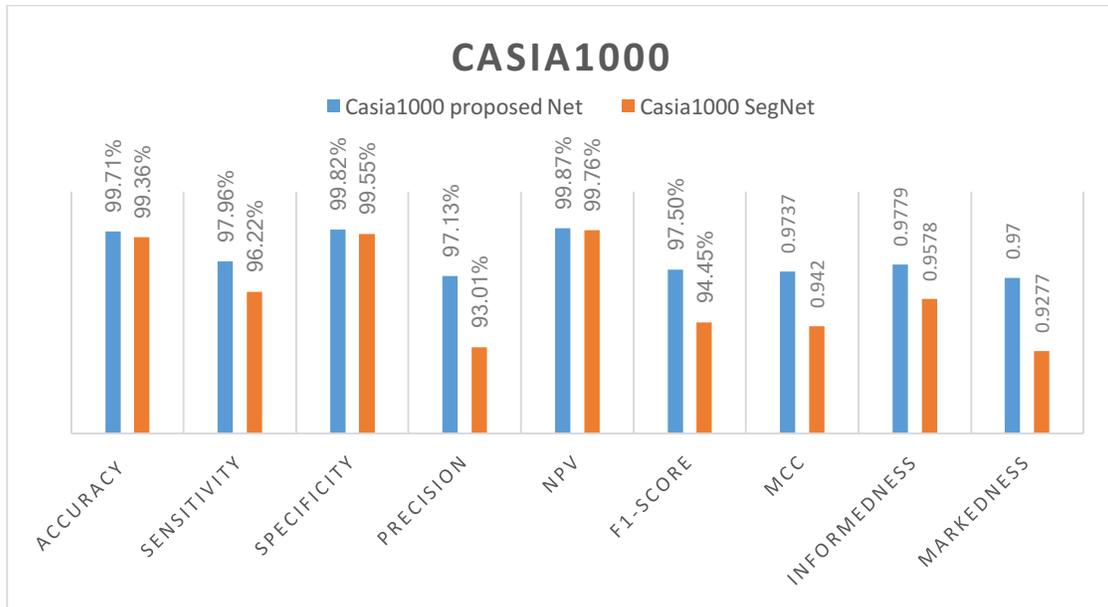

*Figure 22: Comparisons between the presented network and SegNet-basic on CASIA1000. Higher value indicates better performance.*

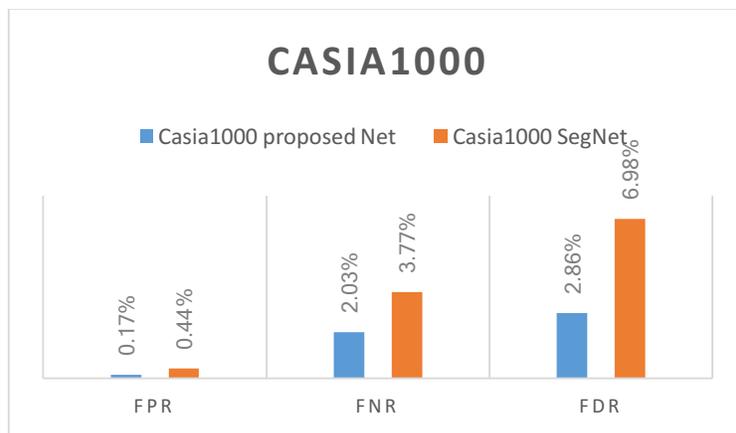

*Figure 23: Comparisons between the presented network and SegNet-basic on Bath800. Lower value indicates better performance.*

From these figures, it is concluded that the proposed method is giving better results on the test set of Bath800 and CASIA1000. Since these two datasets have been used to train both networks, these comparisons show the higher capacity of the proposed method in learning the data distribution in training stage. At the same time, one can comment that learning the training distribution can be a sign of overfitting. In order to investigate this effect, both networks are tested on two other databases UBIRIS and MobBio explained in the next section.

### 5.3.2   Comparing results on UBIRIS and MobBio

It is always important to investigate the performance of a model over databases which hasn't been used in the training stage in order to get better ideas on the model quality in wild environments. In fact, the network is learning the samples which are present in the training set, and a good model should be able to generalize the results for other samples specially unconstrained, consumer graded and difficult ones. In this section, the results for the SegNet-basic network is compared with the proposed network for UBIRIS and MobBio databases (which are not used in the training stage). The results are shown in Figure 24 and Figure 25 for UBIRIS and Figure 26 and Figure 27 for MobBio.

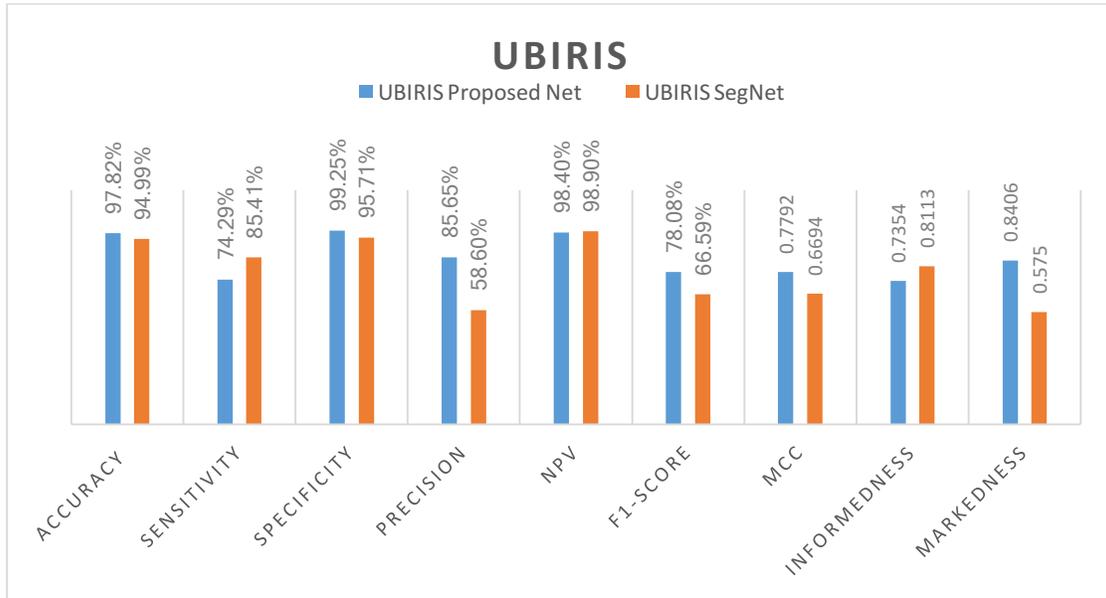

*Figure 24: Comparisons between the presented network and SegNet-basic on UBIRIS. A Higher value indicates better performance.*

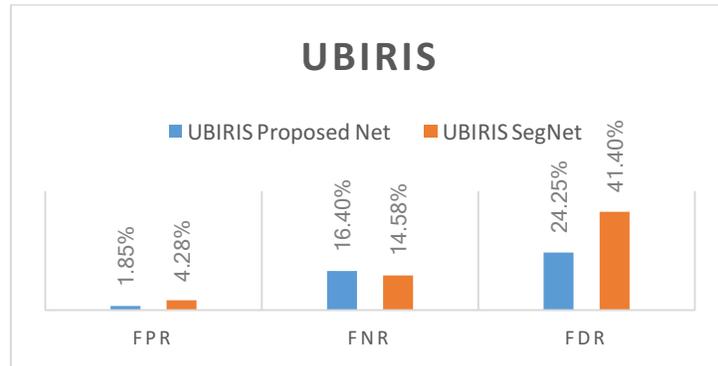

*Figure 25: Comparisons between the presented network and SegNet-basic on UBIRIS. Lower value indicates better performance.*

The presented method is providing higher accuracy than SegNet-basic which implies the better quality in returning true results in the space of all results. This is while SegNet-basic has higher sensitivity and NPV which it means that this architecture is more efficient in ruling out non-iris pixels while the presented model has better performance in finding positive samples due to its higher specificity and precision. Lower FPR shows that the proposed model had lower probability in returning a negative decision and higher FNR shows that SegNet-basic is less probable in making a mistake in returning positive decisions. However, in average the proposed method is more efficient since it has a higher value for F1-score which is the harmonic average of precision and sensitivity. Moreover, also higher MCC shows that the overall performance of the presented network is better than SegNet-basic for UBIRIS dataset.

The numerical results of testing both networks on MobBio dataset are shown in Figure 26 and Figure 27.

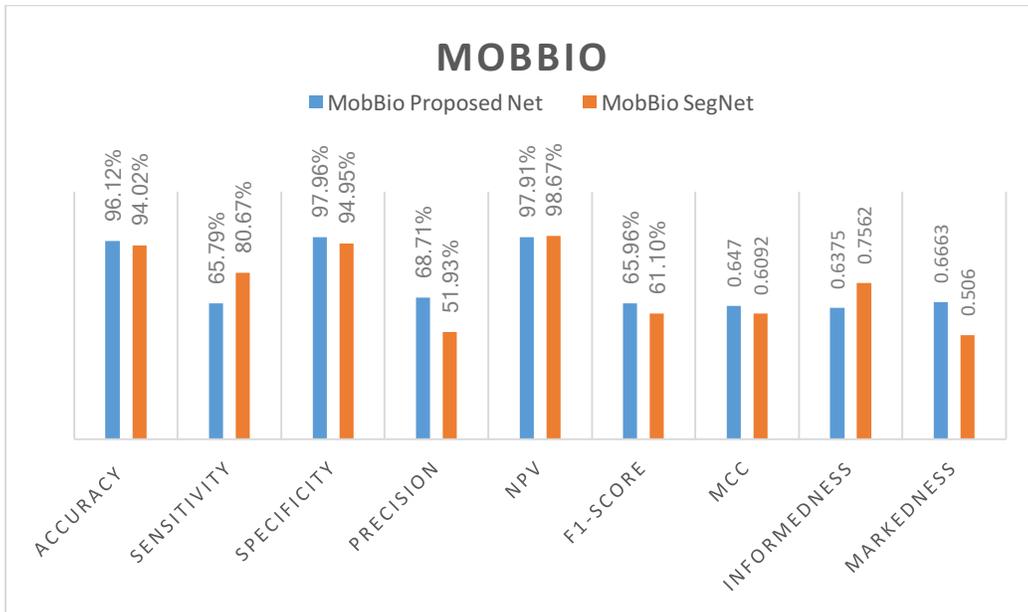

*Figure 26: Comparisons between the presented network and SegNet-basic on MobBio. A Higher value indicates better performance.*

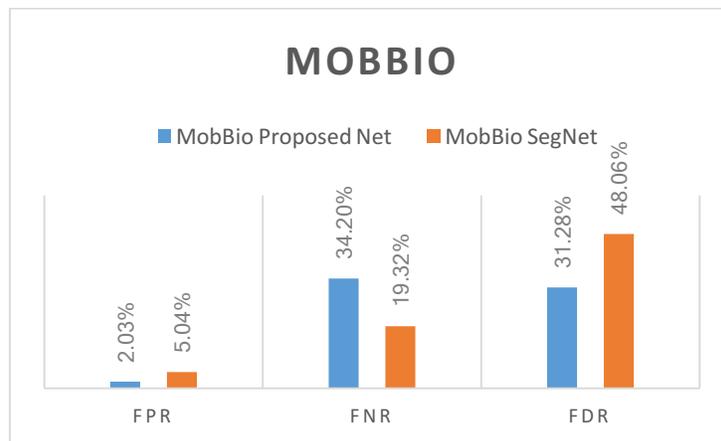

*Figure 27: Comparisons between the presented network and SegNet-basic on MobBio. Lower value indicates better performance*

Results for MobBio database is in the same direction as UBIRIS. The proposed model got higher accuracy which means it has better performance in finding iris pixels. Values for sensitivity and NPV shows better performance of SegNet-basic in ruling out non-iris pixels. And observations on specificity and precision shows the better performance of the proposed model in returning iris pixels. Again the value of FPR shows that the proposed model had lower probability in returning a non-iris pixel and higher FNR shows that SegNet-basic is less likely to make a mistake in returning iris pixels. However, on average the proposed method has better performance due to higher numerical values for F1-score and MCC.

The results of testing both networks on UBIRIS and MobBio datasets demonstrate the overall improved performance of proposed network over SegNet-basic. This shows that the model is not only more capable of learning the training data distribution but also it has a better ability to generalize to unconstrained, wild environments.

## 5.4 Comparison to state of the art

In this next experiment, the proposed method is compared to the most advanced and state of the art segmentation methods in the literature. In the first part, the accuracy of the proposed method is compared to other methods over UBIRIS database. Moreover, in the second part, the sensitivity, precision, and F1-score of the proposed method is compared with some other methods over UBIRIS, MobBio and CASIA databases. The results presented here are the best results of our networks after tuning.

### 5.4.1 Accuracy on the UBIRIS database.

The comparisons of the proposed method with the state of art methods over UBIRIS database is illustrated in Figure 28.

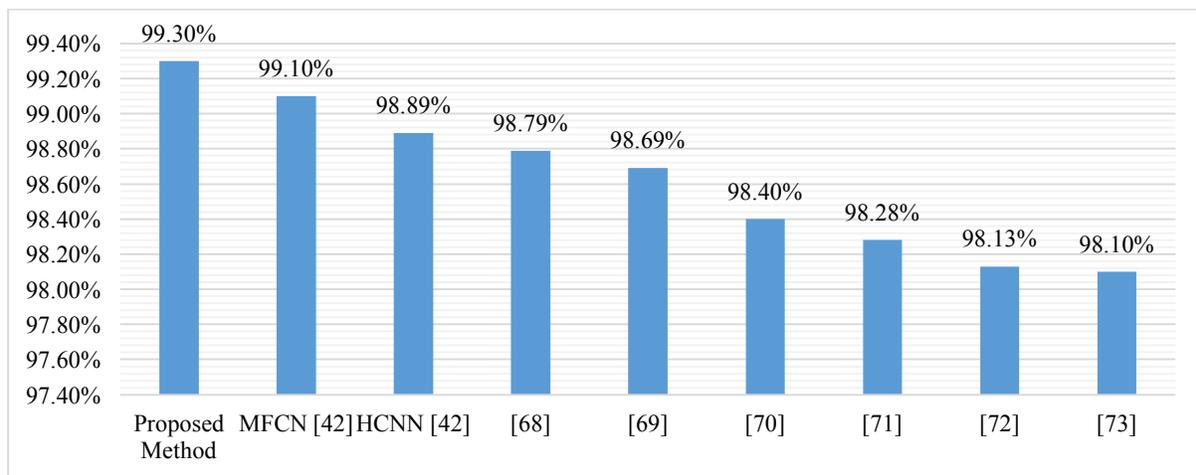

*Figure 28: Accuracy of proposed method vs. other methods over UBIRIS database.*

The MFCN and HCNN [42] methods are using a big, 22 layer, deep neural network to perform the iris segmentation. [68] utilizes the Total Variation (TV) model to overecome the problem of low contrast and noise interference in the eye socket image. [69] proposes an integrodifferential constellation followed by a curvature fitting model to find the iris area. In [70] The Histogram of Oriented Gradients (HOG) is introduced as feature and Support Vector Machine (SVM) is used to perform the automatic segmentation of iris. The random walker algorithm is used to generate the iris map in [71]. In [72] the sclera and iris regions are detected separately using neural networks as classifiers and polynomial fitting is applied estimating the final iris region. [73] proposes a post-classification procedure including reflection and shadow removal and several refinements on pupil and eyelid localizations to get higher performance on iris segmentation task. From Figure 28, the proposed methods gives the best accuracy for UBIRIS database. The main advantage of our work was by selecting the proper data augmentation described in section 2.2. Augmentation step is essential in any model which is designed to work in wild conditions.

The network is learning the distribution of the train set and therefore, designing a set which is representative of the wild unconstraint conditions is crucial in order to get reasonably high performance, i.e., if one can mimic the real-life situations by introducing enough variations to the training set, it is highly probable that the network is able to generalize the learning into non constrained input test samples. Moreover, also tuning is an essential part of our approach to getting a better result for a pre-defined condition. The original network was trained on Bath800 and CASIA1000 databases which are NIR iris images. At the same time, the UBIRIS database is not big enough to train a DNN without encountering over-fitting condition. One of the best approached to train a network on such a small database is to transfer the information from the original network (which is trained on NIR images) and tune it on the new database.

The other advantage of the proposed method is the semi parallel design of the network. In this approach, one can take advantage of several architectures at the same time. This is while the information flow inside the network is not limited to a single path but mixing and merging of several paths.

### 5.4.2 Experiments on sensitivity, precision, and F1-score

The sensitivity, precision, and F1-score of five iris segmentation methods (CAHT, GST, IFFP, Osiris and WAHET) on several databases including UBIRIS, MobBio and CASIA is given in [11]. In our work, the comparisons with these methods have been conducted on the same databases, and the results are given in Figure 29 to Figure 31.

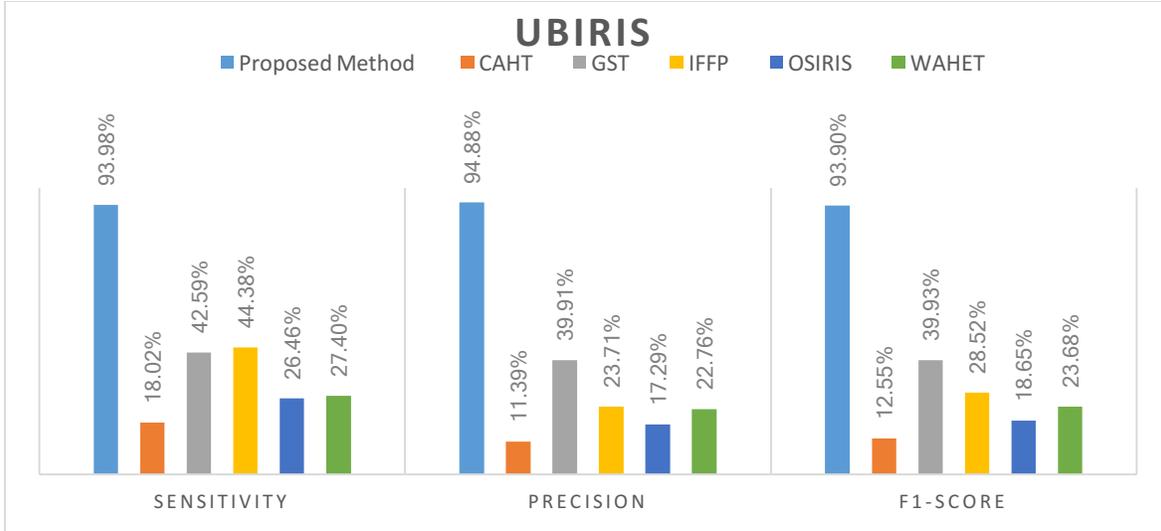

*Figure 29: Sensitivity, Precision and F1-score on UBIRIS database for proposed method vs five other methods. Higher values indicate better performance.*

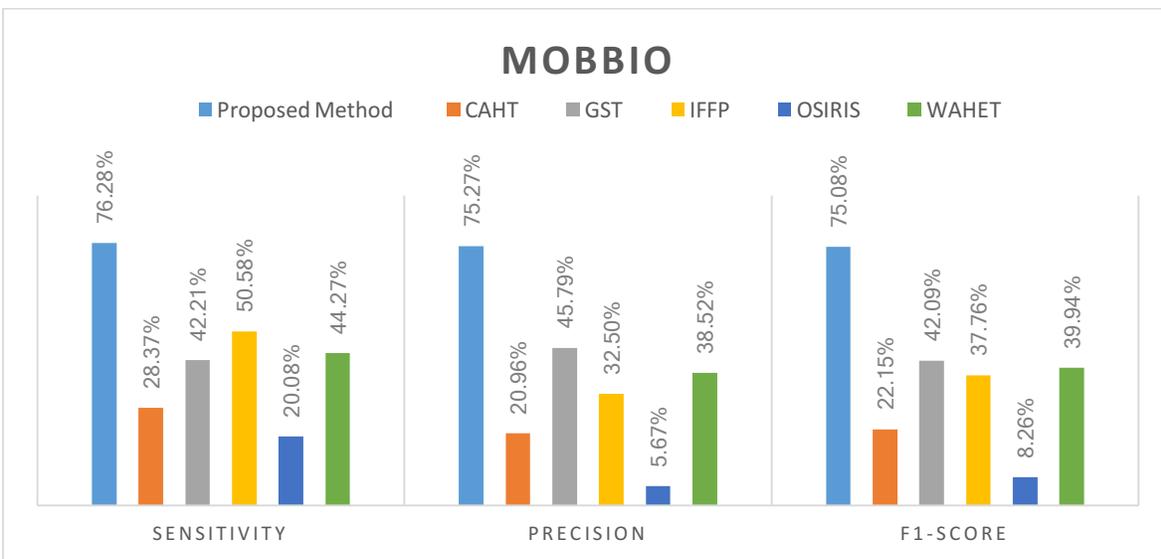

*Figure 30: Sensitivity, Precision and F1-score on MobBio database for proposed method vs five other methods. Higher values indicate better performance.*

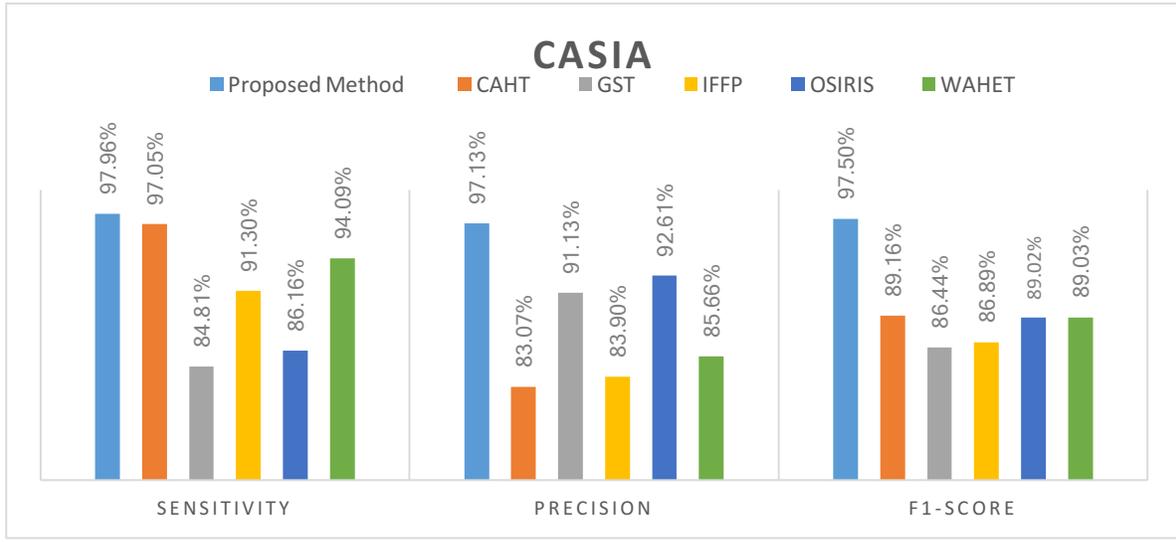

*Figure 31: Sensitivity, Precision and F1-score on CASIA database for proposed method vs five other methods. Higher values indicate better performance.*

Sensitivity and precision metrics measure the quality of the network in ruling out non-iris pixels and detection iris ones respectively; and F1-score is the harmonic average of these two metrics. The higher values correspond to better performance. As shown in Figure 29 to Figure 31, the proposed method gives superior results compared to other approaches on UBIRIS and MobBio databases. Moreover, on the high-quality CASIA database the proposed method is still giving better results. This shows that the proposed method is performing on low-quality consumer graded iris images as good as constrained high-quality samples. This is essential while the user tries to capture the iris information in handheld devices where there is hand shaking, sparse illumination and low-quality front cameras. The proposed network shows that this conditions could be compensated by augmenting the data and also merging several designs into a single network and the numerical results show promising performance of the proposed scheme.

# 6 Conclusions

In this work, a deep neural network framework has been presented to segment the low quality, consumer graded iris images.

There are three main contributions in this work.

i) The data augmentation wherein the high quality eye socket images from Bath800 and CASIA1000 database are degraded and manipulated to give a proper approximation of the low quality images. Four different factors has been considered including image resolution, contrast, shadow and motion blurring. The augmented images give a close approximation of low quality unconstrained iris images.

ii) The recently introduced Semi Parallel Deep Neural Network method has been used to design a fully convolutional network by mixing and merging four parent networks. Each of these networks are taking advantage of different kernel sized and different depths which are extracting and processing different feature levels. The final design is similar to U-Net without pooling.

iii) Inter-database evaluations is giving a more realistic overview of the network performance. Here we can address a very essential problem in deep leaning community wherein the researchers are training a DNN on a specific database and test it on the same database. In this work, every network was tested on Bath800, CASIA1000, UBIRIS, and MobBio. Employing this approach gives a realistic foresight of the performance on real world situations.

The proposed model has been initially trained on the augmented version of the Bath800 and CASIA1000 databases and further experiments were carried out by tuning the original network on UBIRIS and MobBio. Tuned networks were tested on all database and the effect of tuning was widely investigated. Our experiments show that the tuning boosts the performance for the database that the network is tuned on. This is expectable except for databases with very unspecific distributions which will decrease the performance after tuning. Other conclusion is that while designing a model for a specific task, the tuning for that conditions will increase the quality of the model. But if there are not pre-defined situations where the network is implemented, the tuning is not advised.

Since the presented network is a big model which is not easily implementable on a low power handheld hardware, the future works include optimizing the network, training a smaller network, or binarize the model to reduce the calculation and memory usage. Optimizing the network includes reducing the parameter precision down to binary or ternary [74]. This will give, up to 32x memory compression and also reduces the calculation load extensively by eliminating most of multiplications in the model. Other approach is to design a model with smaller number of parameters. Currently, our target is to reduce the number of parameters to the rate of 10x without causing considerable cutback in the performance.

# Appendices

## A: Network Design

The SPDNN method has been used to mix and merge four fully convolutional networks into a single network. These models are shown in Figure 32 to Figure 35.

In [63] the authors used an autoencoder shaped network to perform pixel-wised semantic segmentation. In their model, pooling operation (max-pooling) is applied after convolutional layers in the encoder, and they proposed an un-pooling operation which is performed in the decoder part. In our experiment applying max-pooling after each convolution gave poor segmentation results. We did not use pooling operation in our design and in order to compensate this and to provide the network with the opportunity to cover a significant portion of the input sample with a single filter, larger filter sizes placed in the middle of the network. Getting the output straight from large kernel results in low-quality segmentations accompanied by artifacts, therefore the kernel sizes were gradually decreased to 3x3 in the output layer. Different kernel sizes can handle several feature levels. Bigger the kernel sizes can process coarser features, and smaller kernels are suitable for more detailed features. Figure 32 to Figure 35 show the networks designed for the iris segmentation task. Each of these networks is able to segment the iris images accompanied by artifacts since each design is capable of dealing with just specific levels of features. Smaller networks can detect and process details while ignoring large features. Bigger and deeper networks are taking care of coarser shapes and missing details.

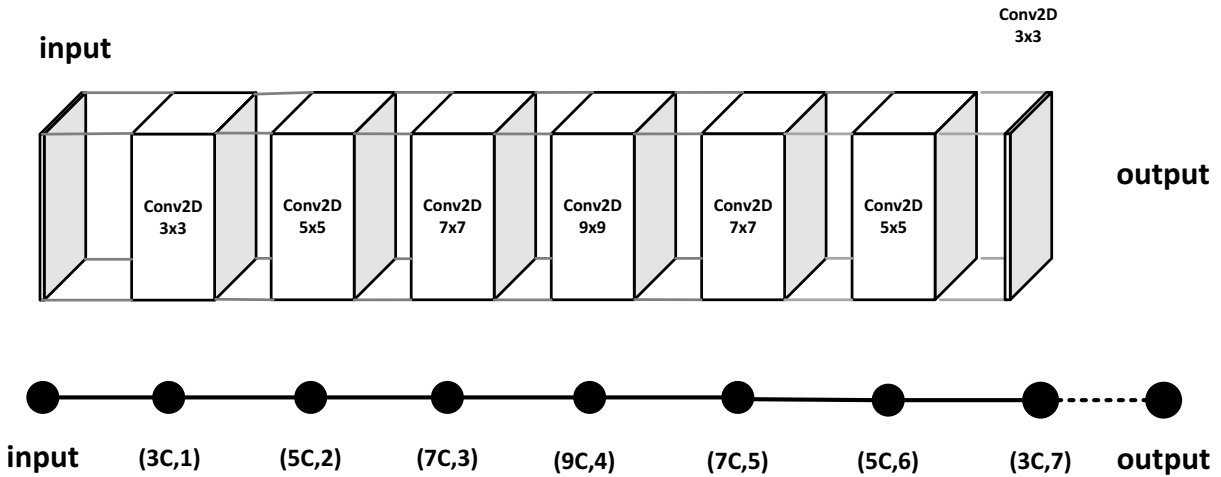

*Figure 32: Up: fully convolutional network designed for iris segmentation task. Down: corresponding graph with labels.*

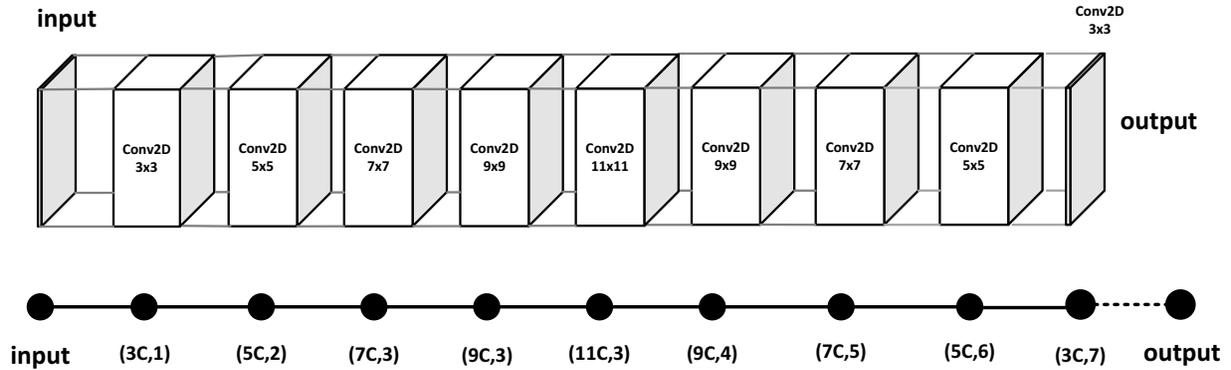

*Figure 33: Up: fully convolutional network designed for iris segmentation task. Down: corresponding graph with labels.*

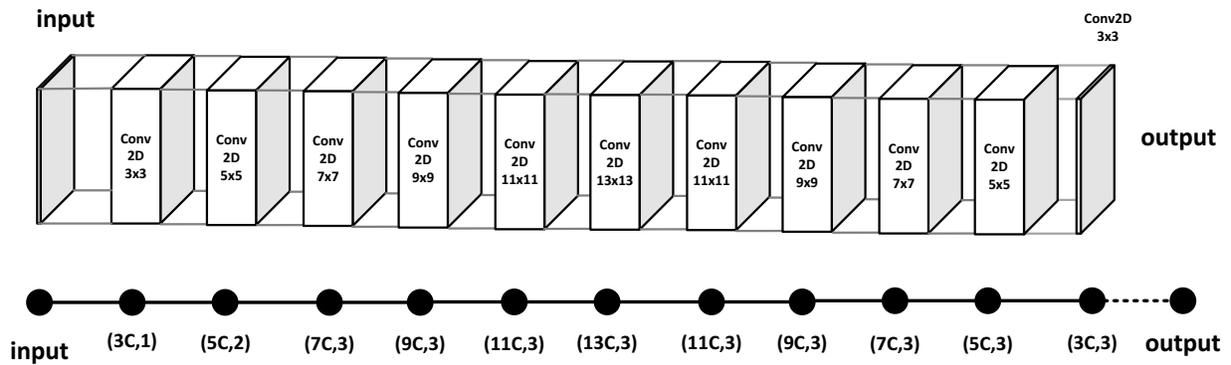

*Figure 34: Up: fully convolutional network designed for iris segmentation task. Down: corresponding graph with labels.*

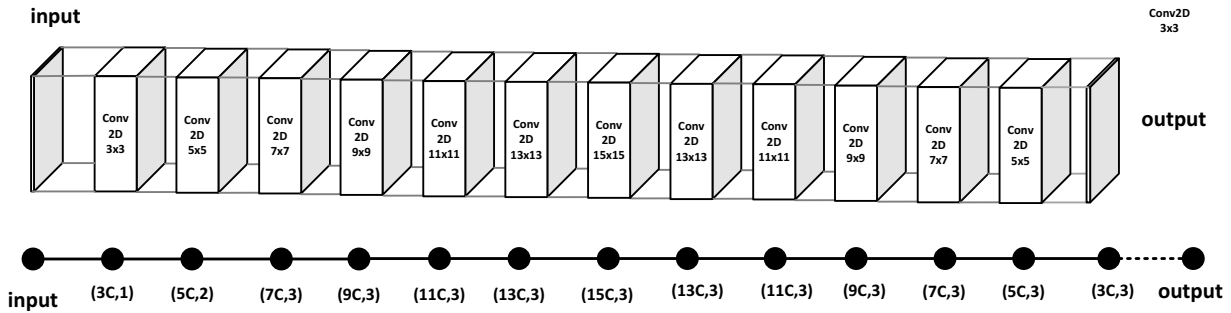

*Figure 35: Up: fully convolutional network designed for iris segmentation task. Down: corresponding graph with labels.*

A.1: Semi Parallel Deep Neural Network v2

The idea of mixing several network architectures in order to take advantage of several designs at the same time is introduced and tested in [48], [49]. In [48], the authors merge eight models into a single architecture by translating deep neural networks into graph space and applying the graph contraction and translate the graph back into a single neural network. Their results show promising improvements in depth estimation using mono camera setup. Moreover, in [49] the convergence and generalization of this approach are discussed.

An improved version of the SPDNN is introduced in this work wherein a second graph contraction is applied to the final graph by introducing a new labeling technique which takes into account the distance of each node from the output. The SPDNNv2 is explained step by step as below:

1. The first step is to translate each network into a graph. Figure 32 to Figure 35 (down) shows the graph correspond to each network. In this method, each layer of the network is considered as a node in the graph.
2. Each node of the graph takes two properties. i) The first property is the layers operation accompanied by the size of the operation. In this labeling method, specific signs are assigned to different operations C for convolutional, F for fully connected layers and P for pooling operation. For example, 7C means a convolutional layer applying a kernel of the size 7x7. There are several rules on labeling fully connected and pooling layers described precisely in [48]. In this work, there are no fully connected layers and pooling operations. ii) The second property assigned to each node is its distance to the input node. For example (7C,3) means that this node corresponds to the third layer which applies 7x7 convolutional operation on its input.
3. The next step is to place all these graphs in parallel, merge input and the output nodes and also devote a letter to each node. The letters assigned to each node based on the properties of the node. The nodes with the same properties get the same letter as their label. See Figure 36. For example, every node with the properties (3C,1) are labeled as A in red and so on.

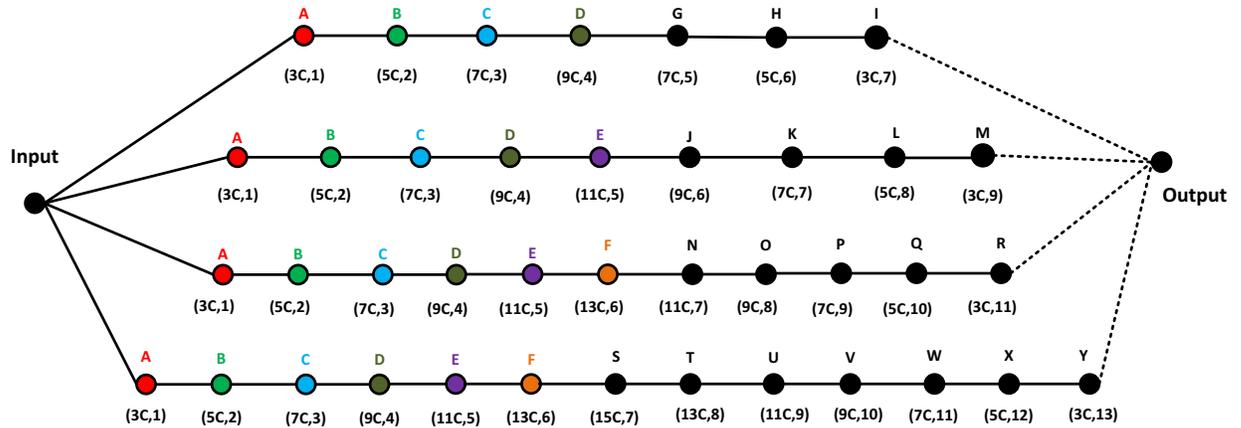

*Figure 36. All the graphs are sharing an input and an output node. Letters are assigned to each node based on their properties. The nodes with the same property get the same label.*

4. In this step, the graph contraction operation is applied to the labeled graph. This will contract the graph by merging nodes with the same label while keeping their connections. For example, the nodes labeled

E will be contracted into a single node while its connections with nodes D, J and F are preserved. The contracted version of Figure 36 is shown in Figure 37.

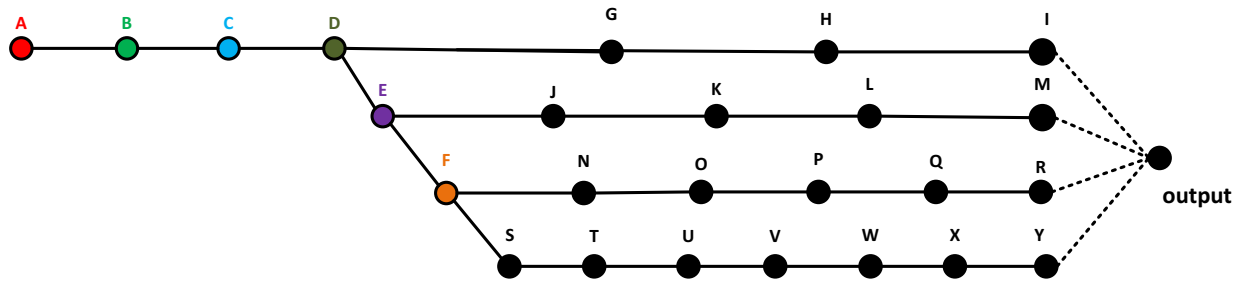

*Figure 37. This figure illustrated the result of graph contraction applied to the graph in **Error! Reference source not found.** Graph contraction is merging the nodes with the same label while preserving its connections.*

5. This step is a complementary to the previous version of the SPDNN method. At this step, a new set of properties are assigned to each node. The first property is borrowed from the previous step which is the operation performed by each node. The second property is the maximum distance from the output to that node. See Figure 38. For example (11C,9) means the node applies a 11x11 convolutional kernel and has a maximum distance of output by nine nodes.

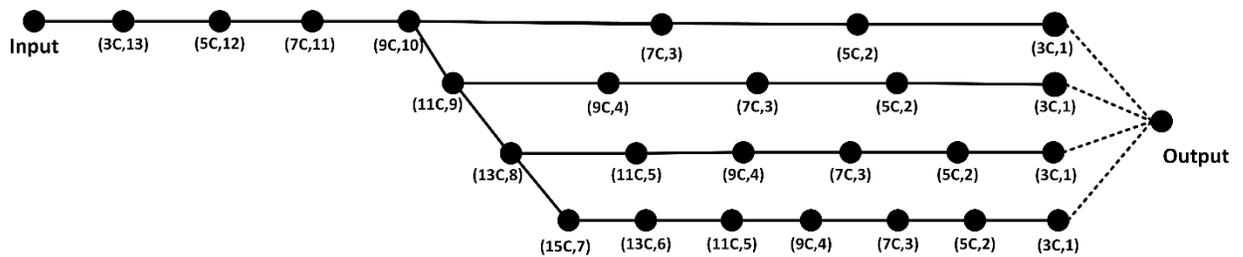

*Figure 38. Anew set of properties are assigned to each node. The first property is the same as previous steps but the second property is the maximum distance of the node to the output node.*

6. This step is similar to labeling presented in step 3. The nodes with the same quality will be assigned the same label. The new labeled graph is shown in Figure 39.

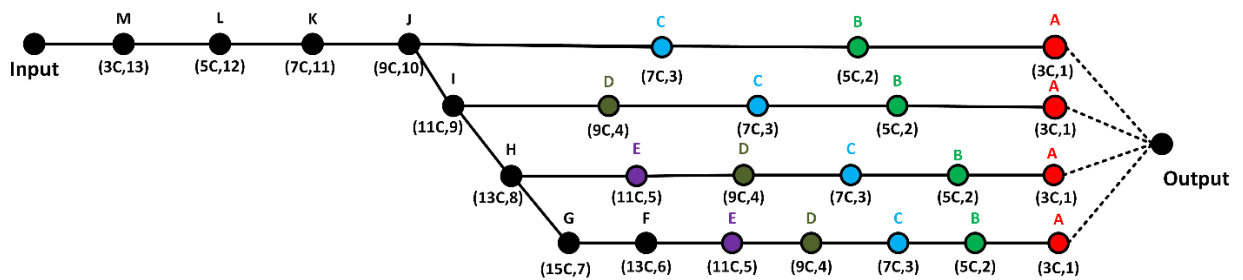

*Figure 39. Nodes with the same properties will be assigned the same label.*

7. Another graph contraction operation is applied to the labeled graph. Merging the nodes with the same label into a single node while preserving their connections to next/previous nodes. The final graph is illustrated in Figure 40.

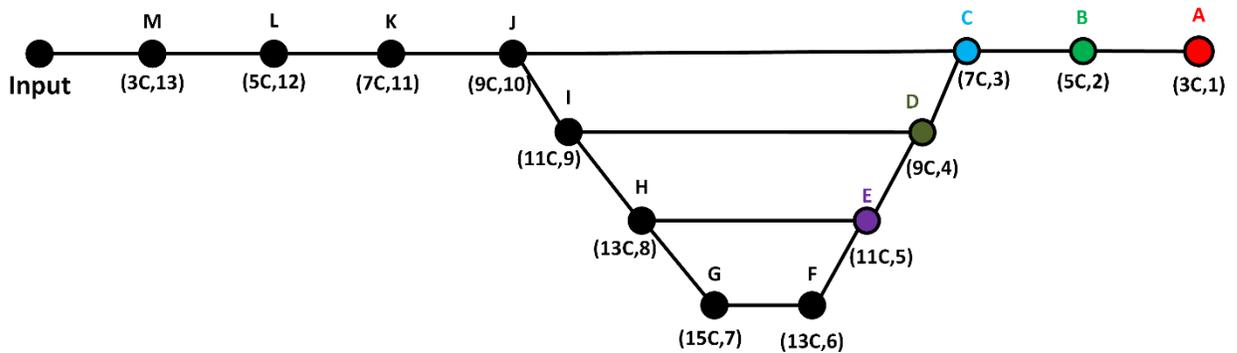

*Figure 40. The graph contraction operation applied again merging the nodes with the same label into a single node.*

8. The last step is to translate the graph back into a neural network. This is mostly done by getting the layer information from the first property of the node. In our work wherever two or more layers are connected to another layer, the concatenation of the layer was fed to the following layer. Figure 41 shows the final network translated back from the graph in Figure 40.

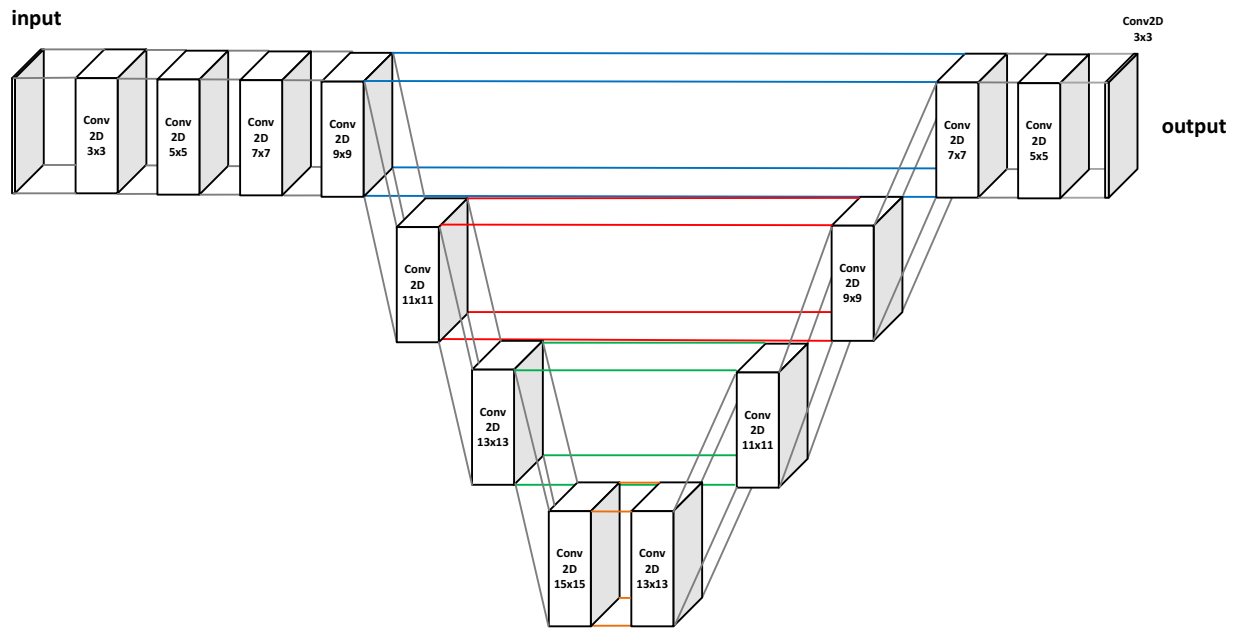

*Figure 41. The network is designed by translating back the final graph and get the layer information from the first property of each node.*

## B: Choosing number of channels in each layer

SegNet [63] is one of the most successful implementations of semantic segmentation using deep neural networks. The fully convolutional architecture of SegNet gives an end to end semantic segmentation model which is one of the most successful segmentation approaches. The original SegNet is made of two VGG16 architecture (just the convolutional layers) placed in a mirrored shape which results in a huge network. The authors in [63] presented an alternative smaller architecture called SegNet-basic. This network is shown in Figure 42. It contains eight convolutional layers. Pooling operation is applied after each layer in encoder part, and the particular un-pooling operation explained in [63] is used in decoder part. Batch normalization is applied after each convolutional layer except the output layer. In order to compare our presented model with SegNet-basic architecture, we trained this architecture given the same training/validation dataset and same training technique. Moreover, for the comparisons to be fair our network is designed in such way to have almost same number of parameters as SegNet-basic.

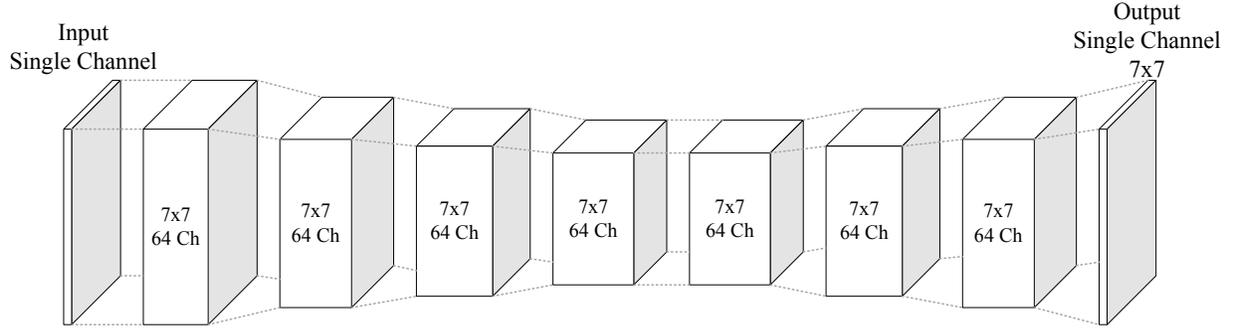

*Figure 42. SegNet-basic contains eight convolutional layers*

The number of parameters for the operation, mapping a layer with $Ch_i$ channels into a layer with $Ch_o$ channels using a $(k \times k)$ kernel is given by:

$$F_{map}(Ch_i, Ch_o, k) = Ch_i \times Ch_o \times k^2 \quad (6)$$

Number of parameters for the SegNet-basic is given by

$$NP_{SegNet-basic} = F_{map}(1,64,7) + F_{map}(64,1,7) + 6 \times F_{map}(64,64,7) \quad (7)$$

$$NP_{SegNet-basic} = 1210496 \quad (8)$$

In the proposed design, we used a larger number of channels for bigger kernels. This came from the idea that the larger kernels are more representative of the problem since they contain more parameters in each layer and so having more channels is giving more opportunity to these kernels in extracting useful features. 3x3 and 5x5 kernels are assigned to have $Ch_p$ channels, 7x7 and 9x9 having $2Ch_p$, 11x11 and 13x13 to have $3Ch_p$, and 15x15 kernel is taking advantage of using $4Ch_p$ channels. The number of the parameters for the proposed network (Figure 41) is given by

$$NP_{presented} = F_{map}(1, Ch_p, 3) + F_{map}(Ch_p, Ch_p, 5) + F_{map}(Ch_p, 2Ch_p, 7) + F_{map}(2Ch_p, 2Ch_p, 9) +$$
$$F_{map}(2Ch_p, 3Ch_p, 11) + F_{map}(3Ch_p, 3Ch_p, 13) + F_{map}(3Ch_p, 4Ch_p, 15) + F_{map}(4Ch_p, 3Ch_p, 13) +$$
$$F_{map}\big((3+3)Ch_p, 3Ch_p, 11\big) + F_{map}\big((3+3)Ch_p, 2Ch_p, 9\big) + F_{map}\big((2+2)Ch_p, 2Ch_p, 7\big) +$$
$$F_{map}(2Ch_p, Ch_p, 5) + F_{map}(Ch_p, 1,3) \quad (9)$$

which leads to

$$NP_{presented} = 11014Ch_p^2 + 18Ch_p \quad (10)$$

In order to have the same number of parameters for presented and SegNet-basic the following equation should hold

$$NP_{presented} = NP_{SegNet-basic} \quad (11)$$

so we have

$$11014 Ch_p^2 + 18 Ch_p = 1210496 \quad (12)$$

solving this equation for $Ch_p$ we have

$$Ch_p = \pm 10.4836 \quad (13)$$

The $Ch_p = 10$ has been used for our design. So the layers with 3x3, 5x5, 7x7, 9x9, 11x11, 13x13, 15x15 kernels are assigned to get 10, 10, 20, 20, 30, 30, and 40 channels respectively.

### C: Tuning; Network experiment

In this experiment, the effect of tuning the proposed network on UBIRIS and MobBio datasets is evaluated. Tuning a pre-trained network is a smart way to adapt a model to new databases without spending too many resources and computational time. Tuning a network is a way of transferring the information from other solutions into a new problem. The tuning procedure is explained in section 3.2, and here the results are discussed.

### C.1: Tuning on UBIRIS

The proposed network trained on Bath800 and CASIA1000 now is tuned on UBIRIS dataset. Both former databases are constructed under NIR illumination. UBIRIS is a visible database, and to get more accurate segmentations on visible sets, one needs to train a network from scratch or tune the pre-trained network on a visible database. Since UBIRIS is small, the tuning is more practical, and also we already trained a network on NIR images which now it is easy to transfer the information from the trained network and just tune the parameters using UBIRIS samples. The tuned network has been tested on all databases, and the results are given in Table 6 and Table 7. Note that the tuned network has been tested on the test set of Bath800, CASIA1000, UBIRIS and all samples of MobBio.

*Table 6: Proposed network trained on Bath800 and CASIA1000, Tuned on UBIRIS. Metrics measured for different databases. Green means better performance and red declares lower quality results. A Higher value of μ and lower value for σ is desirable.*

|  |  | Bath800 | CASIA1000 | UBIRIS | MobBio |
|---|---|---|---|---|---|
| Accuracy | μ | 92.29% | 99.00% | 99.30% | 97.07% |
|  | σ | 7.63% | 0.69% | 0.54% | 3.07% |
| Sensitivity | μ | 68.35% | 93.11% | 93.98% | 76.28% |
|  | σ | 29.34% | 7.77% | 9.45% | 23.43% |
| Specificity | μ | 97.24% | 99.36% | 99.62% | 98.38% |
|  | σ | 3.04% | 0.44% | 0.48% | 1.83% |
| Precision | μ | 81.30% | 89.84% | 94.88% | 75.27% |
|  | σ | 24.10% | 7.63% | 5.40% | 24.02% |
| NPV | μ | 93.59% | 99.58% | 99.60% | 98.49% |
|  | σ | 7.01% | 0.49% | 0.30% | 1.75% |
| F1-Score | μ | 72.67% | 91.27% | 93.90% | 75.08% |
|  | σ | 27.20% | 7.03% | 9.70% | 23.39% |
| MCC | μ | 0.697 | 90.86% | 0.9442 | 0.7398 |
|  | σ | 0.2881 | 7.18% | 0.038 | 0.2448 |
| Informedness | μ | 0.656 | 92.48% | 0.936 | 0.7466 |
|  | σ | 0.302 | 7.90% | 0.0943 | 0.2475 |
| Markedness | μ | 0.749 | 89.42% | 0.9449 | 0.7376 |
|  | σ | 0.2922 | 7.81% | 0.0536 | 0.253 |

*Table 7: Proposed network trained on Bath800 and CASIA1000, Tuned on UBIRIS. Metrics measured for different databases. Green means better performance and red declares lower quality results. Lower value of μ and σ is desirable.*

|  |  | Bath800 | CASIA1000 | UBIRIS | MobBio |
|---|---|---|---|---|---|
| FPR | μ | 2.75% | 0.63% | 0.37% | 1.61% |
|  | σ | 3.04% | 0.44% | 0.48% | 1.83% |
| FNR | μ | 31.64% | 6.88% | 6.01% | 23.71% |
|  | σ | 29.34% | 7.77% | 9.45% | 23.43% |
| FDR | μ | 18.69% | 10.15% | 5.11% | 24.72% |
|  | σ | 24.10% | 7.63% | 5.40% | 24.02% |

From these values, one can conclude that injecting information from the UBIRIS database into the pre-trained network made a significant change in the network performance for UBIRIS database itself. It also confirms the boosting effect on MobBio database. Since both datasets are visible, this can show that the information in UBIRIS helped the model to generalize the results to other visible databases. At the same time, these results indicate the adverse effect of tuning on Bath800 dataset. Apparently, the information given by UBIRIS dataset to the network were contradicting to what is needed to segment Bath800 samples. This comes from the different setups used to capture data in different databases which changes the data distribution. At the same time, the CASIA1000 is not affected by the tuning process. This gives us the knowledge of how similar the setup and data distributions are in UBIRIS and CASIA1000 databases. In fact, the CASIA1000 is a high quality, constrained database and the tuning did not reduce the performance on this dataset dramatically.

C.2: Tuning on MobBio

UBIRIS is a higher quality database compared to MobBio. MobBio is a very low quality, unconstrained, iris database and figuring out the data distributions is a non-trivial task. In order to get a better view of the influence of this network on tuning process, the initially proposed network which has been trained on Bath800 and CAISA1000 databases has been tuned on MobBio. This network then is tested on all databases. The results are shown in Table 8 and Table 9. Note that the tuned network has been tested on the test set of Bath800, CASIA1000, MobBio and all samples of UBIRIS.

Table 8: Proposed network trained on Bath800 and CASIA1000, Tuned on MobBio. Metrics measured for different databases. Green means better performance and red declares lower quality results. A Higher value of μ and lower value for σ is desirable.

|   |   | Bath800 | CASIA1000 | UBIRIS | MobBio |
|---|---|---|---|---|---|
| Accuracy | μ | 95.55% | 99.30% | 98.19% | 94.72% |
|  | σ | 4.33% | 0.56% | 1.89% | 2.93% |
| Sensitivity | μ | 81.50% | 96.23% | 88.36% | 54.84% |
|  | σ | 16.40% | 3.84% | 15.65% | 25.10% |
| Specificity | μ | 99.09% | 99.50% | 99.08% | 97.08% |
|  | σ | 0.99% | 0.48% | 0.66% | 1.79% |
| Precision | μ | 95.12% | 92.43% | 85.78% | 52.88% |
|  | σ | 5.96% | 6.38% | 10.79% | 23.25% |
| NPV | μ | 95.53% | 99.76% | 98.94% | 97.30% |
|  | σ | 4.90% | 0.36% | 1.89% | 1.78% |
| F1-Score | μ | 86.76% | 94.13% | 86.06% | 51.87% |
|  | σ | 12.41% | 4.38% | 13.11% | 23.28% |
| MCC | μ | 0.8514 | 0.9388 | 0.8569 | 0.499 |
|  | σ | 0.1235 | 0.043 | 0.1232 | 0.2406 |
| Informedness | μ | 0.8059 | 0.9573 | 0.8745 | 0.5193 |
|  | σ | 0.1645 | 0.0385 | 0.157 | 0.2594 |
| Markedness | μ | 0.9065 | 0.922 | 0.8472 | 0.5019 |
|  | σ | 0.0799 | 0.0638 | 0.1078 | 0.2425 |

Table 9: Proposed network trained on Bath800 and CASIA1000, Tuned on MobBio. Metrics measured for different databases. Green means better performance and red declares lower quality results. Lower value of μ and σ is desirable.

|   |   | Bath800 | CASIA1000 | UBIRIS | MobBio |
|---|---|---|---|---|---|
| FPR | μ | 0.91% | 0.49% | 0.91% | 2.91% |
|  | σ | 0.99% | 0.48% | 0.66% | 1.79% |
| FNR | μ | 18.49% | 3.76% | 11.63% | 45.15% |
|  | σ | 16.40% | 3.84% | 15.65% | 25.10% |
| FDR | μ | 4.87% | 7.56% | 14.22% | 47.11% |
|  | σ | 5.96% | 6.38% | 10.79% | 23.25% |

Numerical values of these tables confirm the un-deterministic nature of the MobBio database. It shows that the data distribution in this dataset is highly uncorrelated with other databases and also it is not representative of its own test set. All the metrics for all databases indicate the lower performance of the model after tuning with MobBio. It shows tuning with MobBio injects an amount of uncertainty in the final design which is not productive because it did not even improve the results on MobBio itself. One should note that the standard deviation value for the MobBio test set reduced after tuning with this database which was predictable since after tuning, the network is more robust on the database it has been tuned on.

C.3: Tuning on mixed version of UBIRIS and MobBio

An argument on training and tuning is what if one merges samples of several datasets for train/tune purposes and feed the network with this joint information. Will the model learn extra materials? To see the effect of database merging, the training samples of UBIRIS and MobBio databases are mixed and applied to the original network in the tuning process. The tuned network is tested on the test set of all databases, and the numerical results are shown in Table 10 and Table 11.

Table 10: Proposed network trained on Bath800 and CASIA1000, Tuned on UBIRIS+MobBio. Metrics measured for different databases. Green means better performance and red declares lower quality results. A Higher value of µ and lower value for σ is desirable.

|  |  | Bath800 | CASIA1000 | UBIRIS | MobBio |
|---|---|---|---|---|---|
| Accuracy | $\mu$ | 96.09% | 99.24% | 99.24% | 94.81% |
|  | $\sigma$ | 4.74% | 0.46% | 0.60% | 2.91% |
| Sensitivity | $\mu$ | 87.11% | 95.32% | 94.04% | 56.35% |
|  | $\sigma$ | 16.50% | 4.92% | 9.70% | 24.27% |
| Specificity | $\mu$ | 98.10% | 99.49% | 99.53% | 97.11% |
|  | $\sigma$ | 2.30% | 0.36% | 0.55% | 1.68% |
| Precision | $\mu$ | 91.42% | 92.01% | 93.73% | 53.62% |
|  | $\sigma$ | 11.65% | 5.75% | 6.43% | 23.18% |
| NPV | $\mu$ | 97.07% | 99.71% | 99.62% | 97.37% |
|  | $\sigma$ | 4.56% | 0.34% | 0.31% | 1.79% |
| F1-Score | $\mu$ | 88.55% | 93.50% | 93.33% | 53.46% |
|  | $\sigma$ | 13.85% | 4.46% | 10.10% | 22.60% |
| MCC | $\mu$ | 0.8666 | 0.9319 | 0.9383 | 0.5133 |
|  | $\sigma$ | 0.1521 | 0.0449 | 0.0471 | 0.2368 |
| Informedness | $\mu$ | 0.8521 | 0.9481 | 0.9358 | 0.5346 |
|  | $\sigma$ | 0.1716 | 0.0493 | 0.0971 | 0.2528 |
| Markedness | $\mu$ | 0.8849 | 0.9172 | 0.9336 | 0.5099 |
|  | $\sigma$ | 0.1428 | 0.0578 | 0.0641 | 0.2423 |

Table 11: Proposed network trained on Bath800 and CASIA1000, Tuned on UBIRIS+MobBio. Metrics measured for different databases. Green means better performance and red declares lower quality results. Lower value of μ and σ is desirable

|     |     | Bath800 | CASIA1000 | UBIRIS | MobBio |
|-----|-----|---------|-----------|--------|--------|
| FPR | μ   | 1.89%   | 0.51%     | 0.46%  | 2.88%  |
|     | σ   | 2.30%   | 0.36%     | 0.55%  | 1.68%  |
| FNR | μ   | 12.88%  | 4.68%     | 5.95%  | 43.65% |
|     | σ   | 16.50%  | 4.92%     | 9.70%  | 24.27% |
| FDR | μ   | 8.57%   | 7.98%     | 6.26%  | 46.37% |
|     | σ   | 11.65%  | 5.75%     | 6.43%  | 23.18% |

The numerical results show that mixing two databases for tuning purpose is helping the network to generalize the segmentation task compared to when the network is tuned independently on each database. This results can confirm the conclusions of [67]. Mixing two datasets in training/tuning stage can give better results than independently tunes models even on the datasets the network has been tuned on. Further explanations on Table 10 and Table 11 is given in the Appendix D.

## D: Tuning; Database experiment

In this experiment, the evaluations are given on each database for original and tuned networks. This can give an overview of tuning influence on the network performance and also the effect of mixing databases in tuning stage is discussed.

## D.1: Bath800

The proposed network has been trained on Bath800 and CAISA1000 originally. In order to transfer the information from this trained network and apply it to the visible databases, the network has been tuned on UBIRIS, and MobBio datasets individually and also a merged version of these databases. Here we want to see the effect of tuning, on the test set of the Bath800 database. Figure 43 and Figure 44 show the numerical results of testing on four networks.

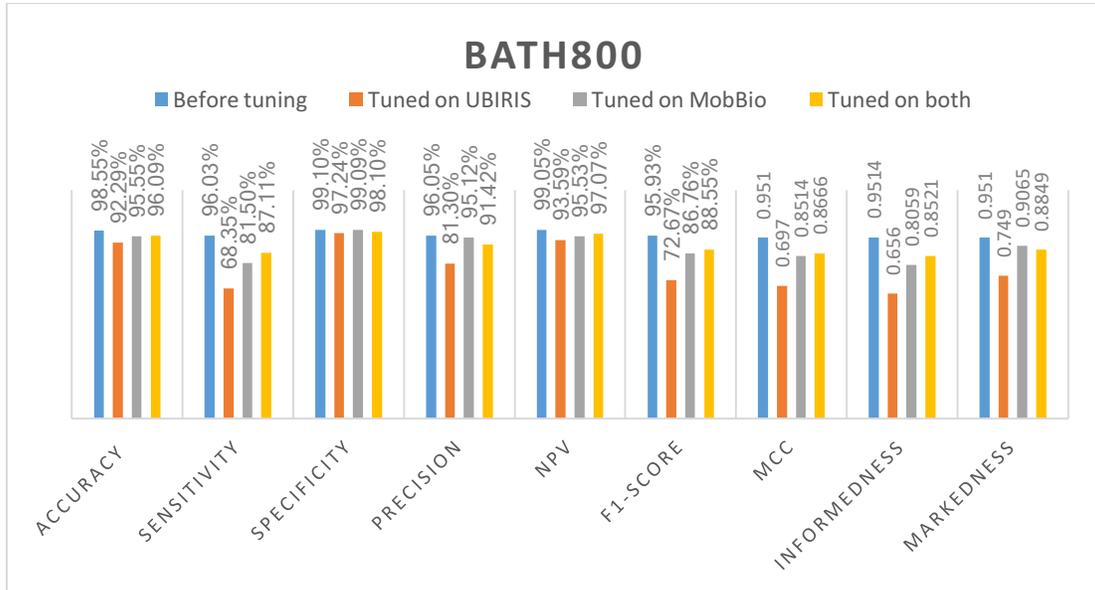

Figure 43: Bath800 tested on four networks before and after tuning. Higher values indicate better performance.

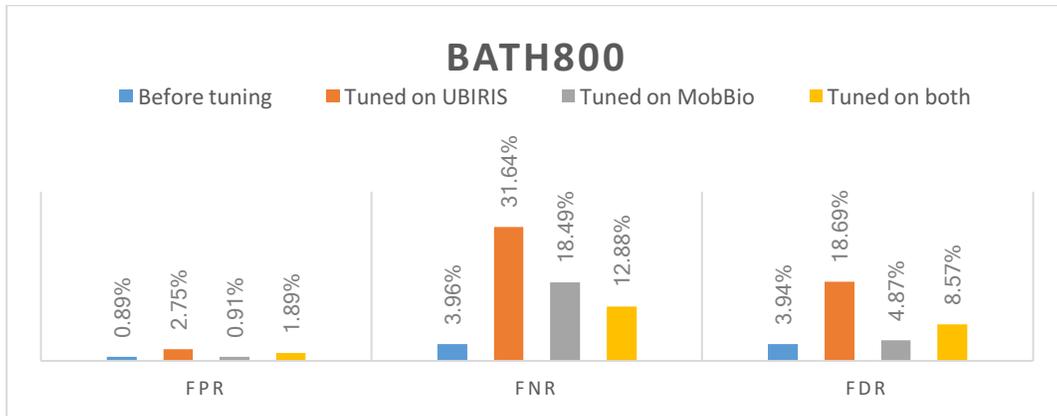

*Figure 44: Bath800 tested on four networks before and after tuning. Lower values indicate better performance.*

All the performances are reduced after tuning the network with any visible database. From these figures, one can conclude that information induced by UBIRIS database was reducing the performance on Bath 800 database more than the information from MobBio did. This can indicate that the kernel used in segmenting iris images in Bath800 is closer to ones needed for MobBio than UBIRIS. The other interesting effect is the results after mixing and merging two databases in tuning stage. The accuracy is slightly better when two databases are mixed compared to tuned on individual ones. Also, sensitivity and NPV are higher, which means that network tuned on mixed data is more efficient in ruling out non-iris pixels compared to networks tuned on individual databases. Since F1-score and MCC are higher as well, one can say the network tuned on both databases is working better on generalizing the results. This shows how important is to mix and merge databases to train/tune a network which is suitable for wild conditions.

D.2: CASIA1000

In this section, the effect of tuning is discussed on CASIA1000 database. Figure 45 and Figure 46 show the numerical results for CASIA1000 tested on four networks.

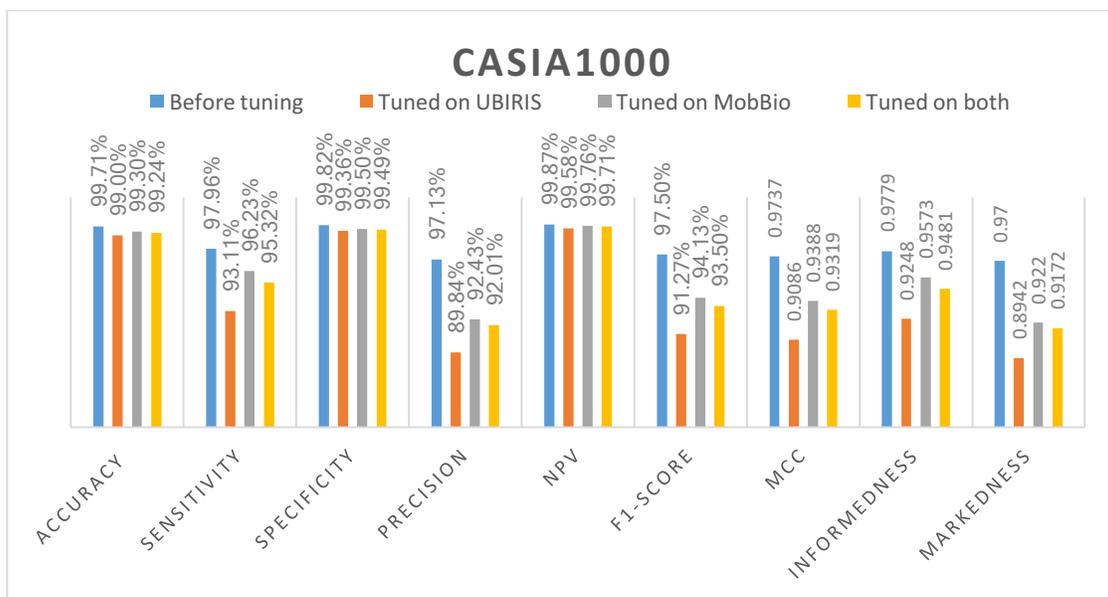

*Figure 45: CAISA1000 tested on four networks before and after tuning. Higher values indicate better performance.*

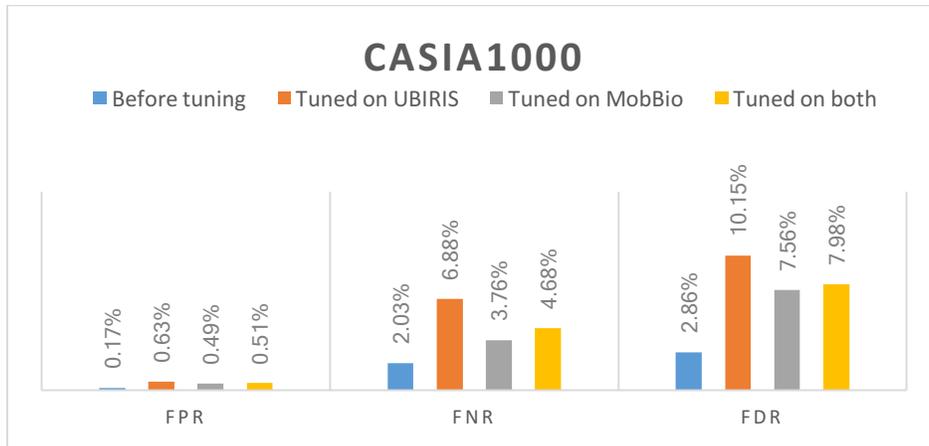

*Figure 46: CAISA1000 tested on four networks before and after tuning. Lower values indicate better performance.*

All the metrics show lower performance on CAISA1000 tuned on any of databases. Numerical results show that tuning on MobBio caused a less negative effect on the performance than other databases. This can have two reasons. The first is the number of samples in MobBio database which lower than UBIRIS and therefore tuning on MobBio alone does not deviate the parameter values compare to UBIRIS database. The other reason is that MobBio is a non-constraint low-quality wild database and tuning on such a database can generalize the network results better than higher quality databases.

### D.3: UBIRIS

In this section, the effect of tuning the network on the UBIRIS database is discussed. Figure 47 and Figure 48, show the numerical results for testing the UBIRIS database on four different networks.

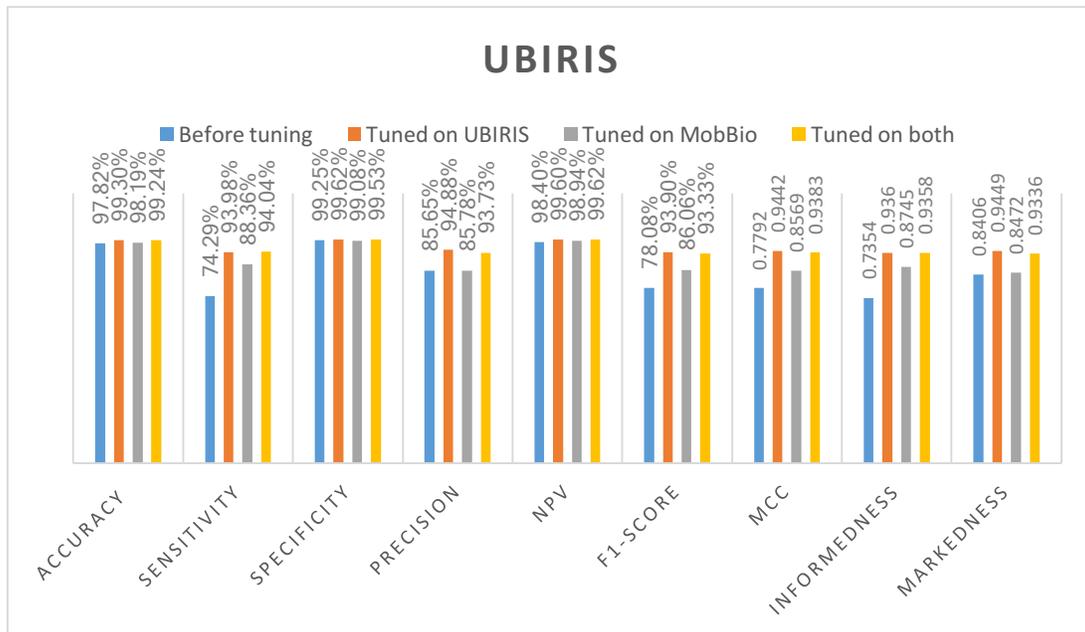

*Figure 47: UBIRIS tested on four networks before and after tuning. Higher values indicate better performance*

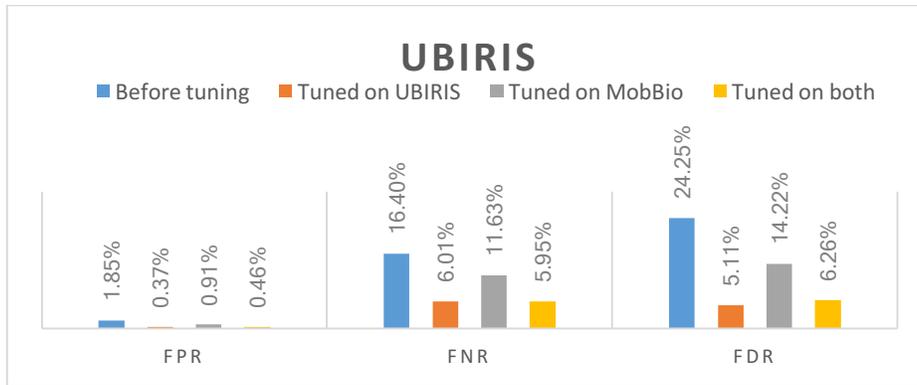

*Figure 48: UBIRIS tested on four networks before and after tuning. Lower values indicate better performance.*

UBIRIS database has better performance on the network which has been tuned on UBIRIS. This database is a medium quality constrained database and tuning the network on it is injecting a set of specific information toward segmenting the samples came from this database. The predictability of the distribution of UBIRIS database can lead the network to gives better results when it is tuned with its samples. By looking at F1-score, MCC, sensitivity, FNR, and FDR, it can be concluded that tuning with MobBio database is increasing the performance on UBIRIS compared to the performance before tuning. This could result from the fact that both databases are visible and have more similar distributions compared to NIR databases.

The results from the network tuned on the combined dataset are close to the network which is tuned on just UBIRIS. This can again show that how mixing and merging several databases can generalize the results.

D.4: MobBio

MobBio dataset is one of the most challenging databases in iris segmentation task. The unconstraint, wild properties of this database makes its result to be highly unpredictable. The numerical results on testing this database on four networks are shown in Figure 49 and Figure 50.

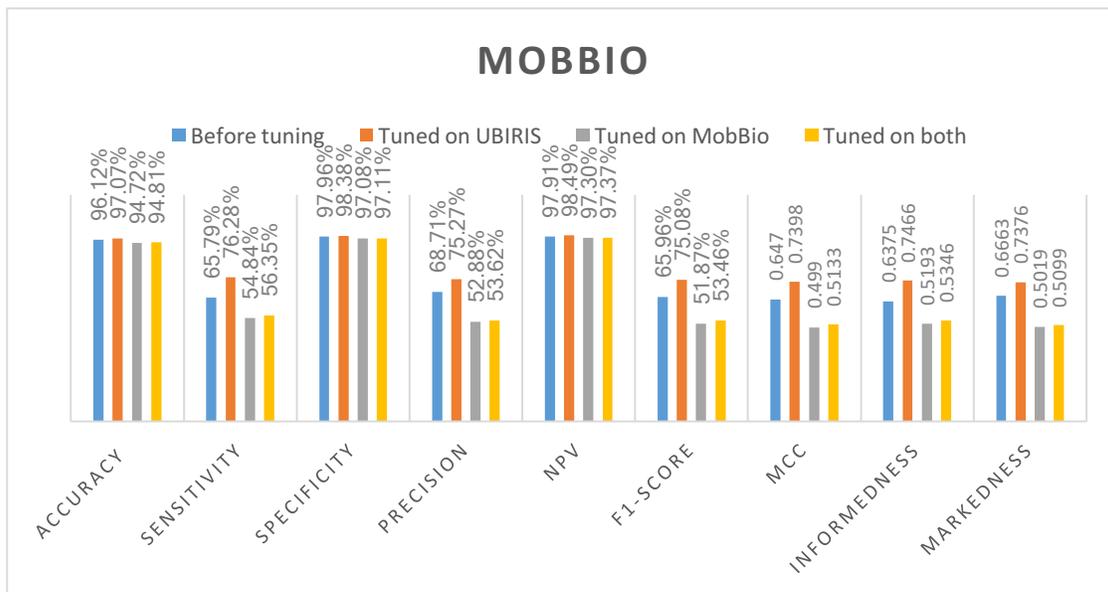

*Figure 49: MobBio tested on four networks before and after tuning. Higher values indicate better performance.*

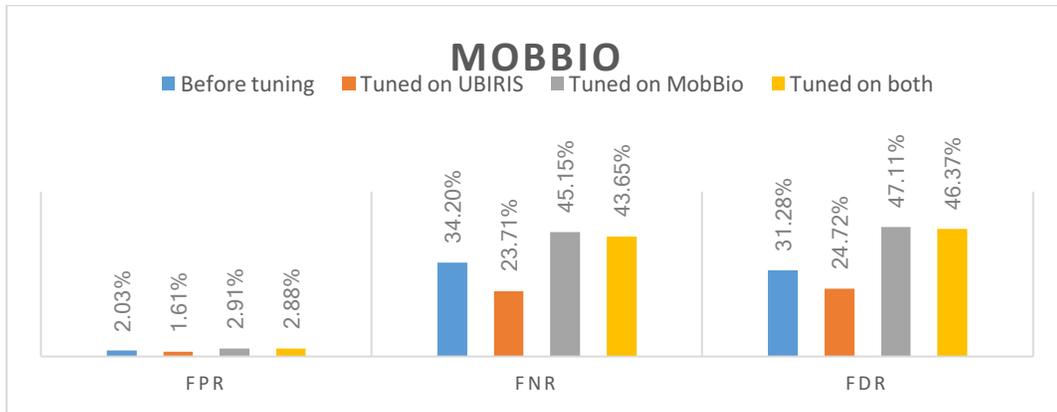

*Figure 50: MobBio tested on four networks before and after tuning. Lower values indicate better performance.*

MobBio is a small database; there are not enough samples to derive a robust distribution of the data while tuning the network with it. It seems that even the test set of MobBio was not representative of its train set. Because MobBio gave better performance on the network that has been tuned on UBIRIS. High variations in the samples of MobBio injects an amount of uncertainty to the network which is useful for generalizing the model, but at the same time, it will reduce the performance on the test set of the same database. UBIRIS is a visible database as well, and apparently, information injected by this database to the network could generalize the network to boost the performance while testing on MobBio. Also, it is shown that while considering the MobBio database, the results on network tuned on both databases is close to the values for the network which has been tuned just on MobBio dataset. It means that the distribution of MobBio samples was dominant in tuning stage although the number of samples of UBIRIS is more than MobBio. It shows that when testing on challenging datasets having a robust distribution is better than high variations in the train set.

D.5: Average results

In this section, the average of metric values for all databases tested on different networks is presented. See Figure 51 and Figure 52.

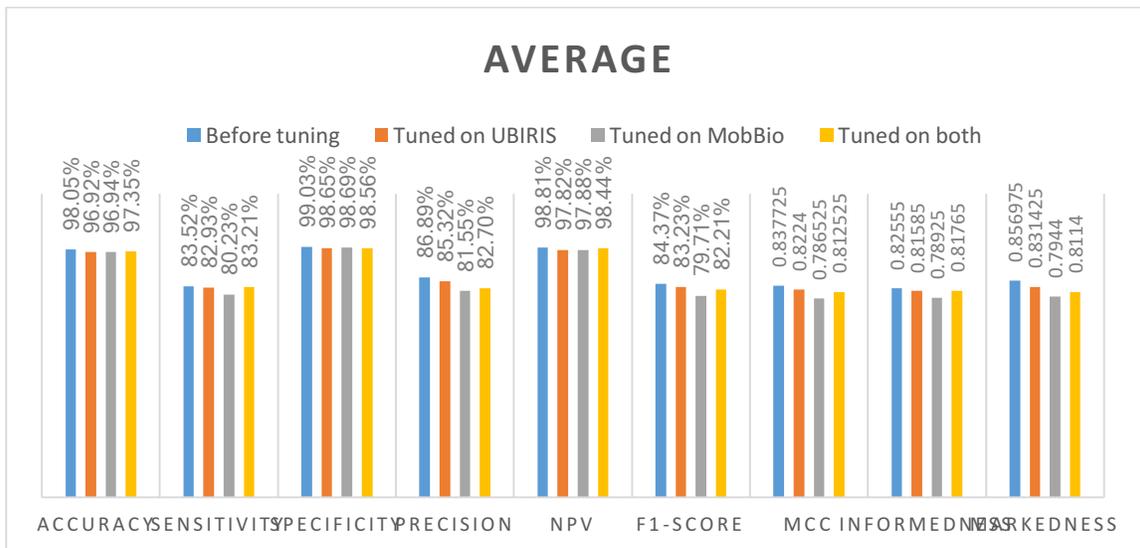

*Figure 51: Average values for all databases tested on four networks before and after tuning. Higher values indicate better performance.*

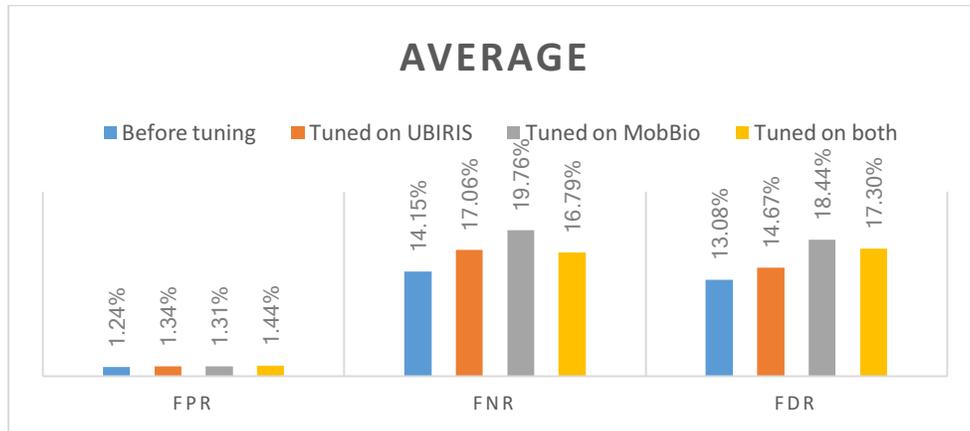

*Figure 52: Average values for all databases tested on four networks before and after tuning. Lower values indicate better performance.*

These observations show better performance for the original network before tuning on any visible database. This means that introducing any new variation to the network is accompanies by injecting more uncertainty to the network which reduces the performance in average. From two previous sections one can say for visible applications, the tuning boosted the performance of the network. It is evident that tuning the network on visible datasets result in a higher quality model while testing on visible samples and vice versa. However, if one needs to have a network for an unknown input (visible or NIR), tuning is not recommended.